\documentclass[12pt]{iopart}

\usepackage{amssymb,amsfonts}
\usepackage{graphicx,times}
\usepackage[authoryear]{natbib}
\begin{document}

\title[]{A computational study of high-speed microdroplet impact 
onto a smooth solid surface}

\author{James Q.~Feng}
%\maketitle

\address{Optomec, Inc., 2575 University Avenue, Suite 135, St. Paul, Minnesota 55114, USA}
\ead{jfeng@optomec.com}

\begin{abstract}
Numerical solutions of high-speed microdroplet impact onto a 
smooth solid surface are computed, using the 
{\em interFoam} VoF solver of the 
OpenFOAM$^{\circledR}$ CFD package.
Toward the solid surface, the liquid microdroplet 
is moving with an impinging gas flow,
simulating the situation of 
ink droplets being deposited onto substrate 
with a collimated mist jet in the Optomec
Aerosol Jet$^{\circledR}$ printing process.
For simplicity and computational efficiency,
axisymmetric incompressible flow is assumed in 
the present study of
the free-surface fluid dynamic problem.
The computed values of maximum spread factor, 
for the range of parameters of practical interest to 
Aerosol Jet$^{\circledR}$ printing,
were found in very good agreement with some of 
the correlation formulas proposed by previous authors in the literature.  
Combining formulas selected from different authors 
with appropriate modifications yields a
maximum spread factor formula that
can be used for first-order evaluations of deposited 
in droplet size during the
Aerosol Jet$^{\circledR}$ technology development.
The computational results also illustrate 
droplet impact dynamics with lamella shape 
evolution throughout the spreading, receding-relaxation, 
and wetting equilibrium phases, consistent with 
that observed and described by many previous authors.
This suggests a scale-invariant nature of the basic droplet impact 
behavior such that experiments with larger droplets at 
the same nondimensional parameter values may be 
considered for studying microdroplet impact dynamics.
Significant free surface oscillations can be observed
when the droplet viscosity is relatively low.
The border line between periodic free surface oscillations and 
aperiodic creeping to capillary equilibrium free surface shape 
appears at the value of Ohnesorge number around $0.25$.
Droplet bouncing after receding is prompted with
large contact angles at solid surface (as consistent with 
findings reported in the literature),
but can be suppressed by increasing the droplet viscosity.

\end{abstract}

\vspace{2pc}

\maketitle

%\newpage

%% main text
\section{Introduction}
\label{}

With the Aerosol Jet$^{\circledR}$ direct-write technology,
ink microdroplets generated by a liquid atomization process
are deposited onto a substrate in 
a form of collimated mist stream 
(which can become less than $10$ $\mu$m in diameter 
having the ink droplet concentration typically about $50$ nL/cc) 
with considerable 
impinging velocity, e.g., 20 to 100 m/s
\citep[cf. ][]{renn2006, zollmer2006,
renn2007, hedges2007, kahn2007,
renn2009, renn2010,
christenson2011, paulsen2012}.
Therefore, the ink droplets can have sufficient momentum
to impact the substrate several millimeters away from
the deposition nozzle as directed by 
the high-speed jet flow \citep[cf. ][]{feng2015}.
The 
Aerosol Jet$^{\circledR}$
functional inks typically consist of suspensions of nano-particles or 
polymer solutions formulated with 
appropriate properties such that they can be adequately
aerosolized with a liquid atomizer.
The ink droplet diameter is usually in a narrow range of $1$ to $5$ $\mu$m
with the volume mean diameter around $2.5$ $\mu$m, 
such that fine features as small as $\sim 10$ $\mu$m 
can be produced by the additive manufacturing process.
As with many industrial applications such as spray coating, 
inkjet printing, and so forth, 
understanding of droplet deposition behavior is 
important for achieving desired  
Aerosol Jet$^{\circledR}$
print quality.
Therefore,
a detailed analysis of 
high-speed microdroplet impact on a solid surface 
can provide practically valuable insights.

The process of droplet impact on a surface 
involves a rich set of free-surface fluid dynamics phenomena,
ranging from spreading, receding, oscillating, to 
bouncing, splashing, etc. \citep{yarin2006}.
It has been a subject of intensive study by many authors
\citep[e.g.,][as well as references cited therein]{ford1967,
foote1974, 
chandra1991, rein1993, healy1996,
bussmann1999, bussmann2000, sikalo2002, rioboo2001, rioboo2002, 
toivakka2003, law2015},
for its relevance to a wide variety of applications. 
Yet, our understanding of the associated fluid dynamices may still
be far from thorough,
probably due to the difficulties in consistent characterizations of 
wetting and surface properties
as well as lack of agreeable formulations of 
moving contact line boundary conditions for theoretical modeling. 
For example, 
numerous empirical and semi-empirical formulas 
were proposed for describing
the maximum spread factor, 
defined as the maximum normalized contact diameter 
of the lamella at the end of 
spreading phase,
for its practical importance
\citep[e.g.,][]{scheller1995,pasandideh1996, 
toivakka2003, attane2007, roisman2009, german2009};
each has an apparently different form and 
quantitative agreement with each other 
for a given case
usually does not seem as good as one would hope 
\citep[cf.][]{perelaer2009, ravi2010, visser2012, visser2015}.
This makes it very difficult to decide which formula to use
for estimating the spot size
as a result of each ink droplet impact on the substrate 
with parameters of particular interest.

Because the 
Aerosol Jet$^{\circledR}$
printing process involves micron-size droplets 
carried by a high-speed impinging gas jet at velocity
typically around $50$ m/s,
experimental investigations can be quite challenging
and prohibitively expensive,
if not impossible.
Rather recently \cite{visser2012, visser2015}
have reported experimental measurements of microdroplet impact
with an interferometic technique 
that enabled sub-micron spatial resolution 
at frame rates exceeding $10^7$ per second,
which still seems to be an order of magnitude short 
for the 
Aerosol Jet$^{\circledR}$
situation.
To date, computational analysis 
with numerical solutions of the governing equations
may be the only option 
for gaining insights into the micron-size droplet 
impact at high velocity 
relevant to the
Aerosol Jet$^{\circledR}$
additive manufacturing process.

Due to extensive free surface deformations involved in
droplet impact process, numerical computations have remained 
challanging.  Although the explicit interface tracking
method with boundary-fitted moving mesh 
using an arbitrary Lagrangian-Eulerian scheme offers 
the highest accuracy for the free-surface flow problem,
it is mostly effective for the types of problems 
with moderate free surface deformations without
topology changes \citep[e.g.,][]{feng2010, feng2015} 
and becomes too complicated to be practically applicable to  
the situation of droplet impact problem where phase topology
can change with significant free surface movement or even 
disintegration.
On the other hand,
the implicit Eulerian interface capturing methods such as volume of fluid (VoF)
have been developed for effective computations of flows
involving substantial topology changes with interface breaking
(because the mesh does not need to move with the interface),
despite some compromise of numerical accuracy.
Among many versions of the VoF solvers, 
the {\em interFoam} of an open-source CFD package called
OpenFOAM$^{\circledR}$ has been attractive to numerous users
and validated by many authors 
\citep{berberovic2009, saha2009, deshpande2012, morgan2013, hoang2013}.
Over years of code development and testing,
the numerical algorithms implemented in {\em interFoam} 
have been continuously improved to enable reasonably accurate interface 
representation with effective advection treatment,
handling large density ratios, 
reducing ``spurious (parasitic) currents'', 
and so forth \citep{gopala2008, deshpande2012}.

The purpose of the present work 
is to compute axisymmetric solutions of 
a droplet impact
on a solid surface 
with parameters relevant to the  
Aerosol Jet$^{\circledR}$
printing,
using the well-established {\em interFoam} 
VoF solver.
Through detailed comparisons of the computed results 
with the available formulas 
from different authors,
a reliable correlation formula for 
the maximum spread factor can be obtained
for predicting the deposition spot size on substrate corresponding to
a single ink droplet impact.
Such spot sizes directly relate to achievable resolutions 
with the ink droplet size distribution
involved in the    
Aerosol Jet$^{\circledR}$
additive manufacturing process.
Moreover, the numerical solutions can also reveal 
other possible outcomes beyond spreading phase, such as 
oscillations, bouncing, etc.
In what follows, 
a brief description of the problem formulation 
and computational method is presented in section 2,
the numerical results for Weber number $We = 2500$,
$100$, and $5$ are presented in section 3 
with discussion of implications for 
Aerosol Jet$^{\circledR}$
printing given in section 4,
and concluding remarks are provided 
in section 5.

\section{Problem statement and computational method}\label{method}
Considered here is a liquid droplet 
of density $\rho_d$, viscosity $\mu_d$, surface tension $\sigma$,
and diameter $d$ 
impacting a smooth solid surface
as carried by a gas impinging flow at velocity $U$ (figure 1).
The surrounding gas has density $\rho_g$ and viscosity $\mu_g$. 
Solutions to
the Navier-Stokes equations for incompressible Newtonian fluids are
computed using the {\em interFoam} VoF solver of the 
OpenFOAM$^{\circledR}$
CFD package. 

\begin{figure}%[htb]
\centering
{\includegraphics[clip=true,scale=0.56,viewport=5 20 650 500]{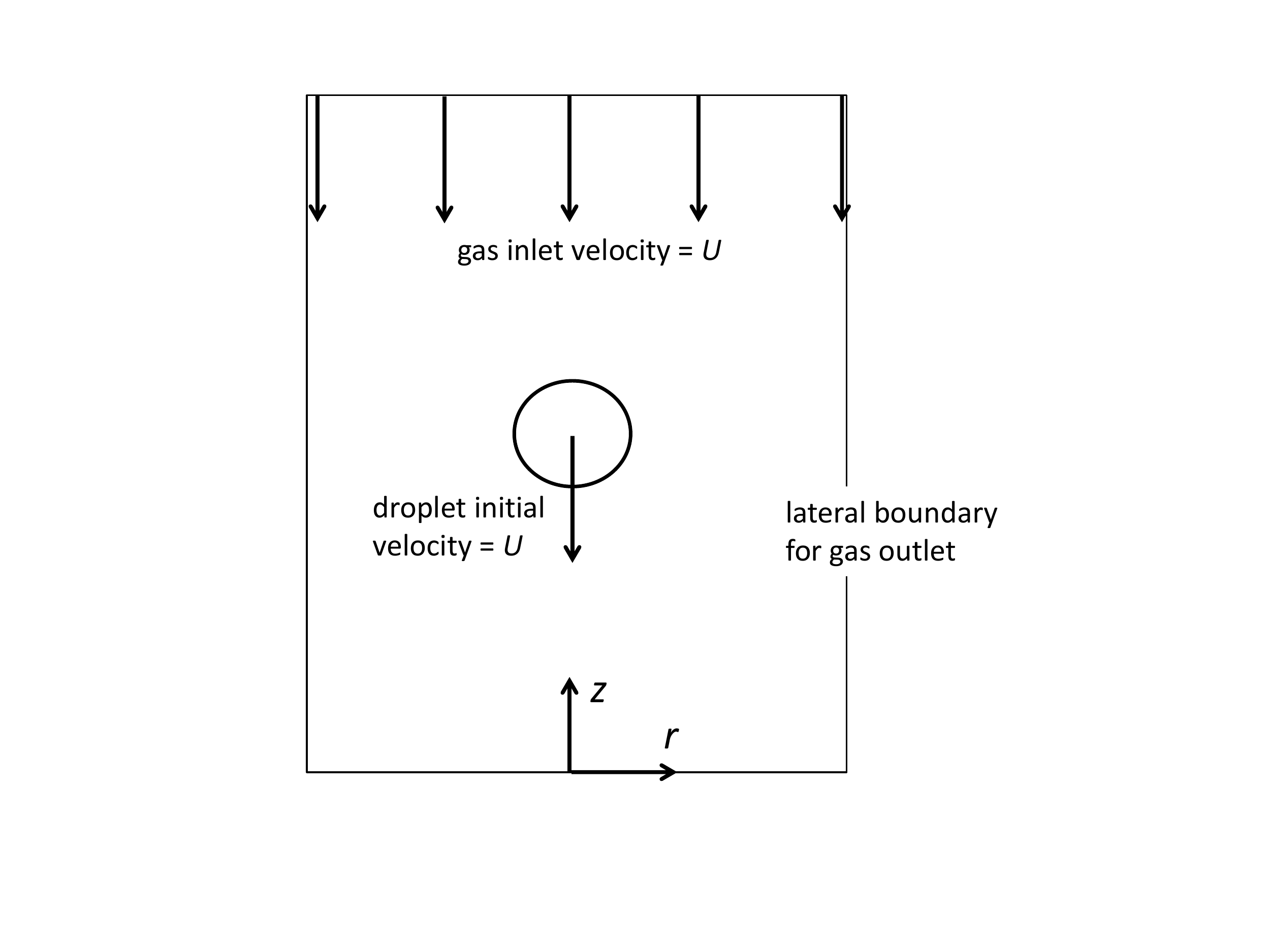}}
\caption{Schematics of a droplet moving with impinging gas  
flow at velocity $U$ to impact a 
solid surface.}
\label{fig:fig1}
\end{figure}

With the volume-of-fluid (VoF) method, an indicator function $\alpha$ 
(also called the phase fraction function) 
is used to represent the volume fraction of one of the phases.
The discontinuity at a gas-liquid interface is represented by
a gradient of the continuous function $\alpha$.
Therefore, the interface is rendered as a diffuse layer 
with finite thickness on the order of the finite volume cell size,
and the interface location may not be determined 
precisely with sub-grid resolution.
Two immiscible fluids are treated as one effective fluid throughout 
the problem domain, having a continuously distributed 
phase fraction function $\alpha$
($0 \le \alpha \le 1$) 
as well as distributed density $\rho$ and viscosity $\mu$ according to
\begin{equation}\label{rho-mu}
\rho \equiv \rho_d \alpha + \rho_g (1 - \alpha) \mbox{ and } \,
\mu \equiv \mu_d \alpha + \mu_g (1 - \alpha) \,
\quad .
\end{equation}
To improve the interface resolution, 
the transport equation for phase fraction function 
used in {\em interFoam} is of the form 
\citep{ubbink02, rusche2002, berberovic2009}
\begin{equation}\label{phase-fraction}
\frac{\partial \alpha}{\partial t}
+\nabla \mathbf{\cdot} (\alpha \mathbf{u})
+\nabla \mathbf{\cdot} [\alpha (1 - \alpha) \mathbf{u}_r] = 0
\quad ,
\end{equation}
where the velocity of the effective fluid and relative velocity are
respectively
\[
\mathbf{u} \equiv \alpha \mathbf{u}_d + (1 - \alpha) \mathbf{u}_g \,
\mbox{ and } \,
\mathbf{u}_r \equiv \mathbf{u}_d - \mathbf{u}_g \,\, .
\]
Then, the momentum equation can be written as
\begin{equation}\label{momentum}
\frac{\partial (\rho \mathbf{u})}{\partial t}
+\nabla \mathbf{\cdot} (\rho \mathbf{u} \mathbf{u}) 
- \nabla \mathbf{\cdot} (\mu \nabla \mathbf{u})
- (\nabla \mathbf{u}) \mathbf{\cdot} \nabla \mu
= - \nabla p - g \, \mathbf{e_g} \mathbf{\cdot} \mathbf{x} \nabla \rho 
+ \sigma \kappa \nabla \alpha  \,\, , 
\end{equation}
where $\sigma$ denotes the interfacial tension and $\kappa$ the mean
curvature of the free interface, determined from 
\begin{equation}\label{mean-curvature}
\kappa \equiv - \nabla \mathbf{\cdot} \left(\frac{\nabla \alpha}
{|\nabla \alpha|}\right) \,
\quad .
\end{equation}
In (\ref{momentum}), $g$ is the value and 
$\mathbf{e_g}$ 
the unit vector of gravitational acceleration,
$\mathbf{x}$ is the position vector, 
and $p$ the lumped (or piezometric) pressure
defined as
\[
p \equiv p_0 - \rho g \, \mathbf{e_g} \mathbf{\cdot} \mathbf{x} \, ,
\]
with $p_0$ denoting the thermodynamic pressure.
For incompressible flow, the velocity field also 
satisfies the continuity equation
\begin{equation}\label{divergence-free}
\nabla \mathbf{\cdot} \mathbf{u} = 0
\quad .
\end{equation}

As illustrated in figure 1,
no-slip boundary condition ($\mathbf{u} = \mathbf{0}$) is applied 
at the solid surface ($z = 0$),
and a uniform-fixed-value velocity ($\mathbf{u} = -U \mathbf{e_z}$) 
is specified 
at the inlet ($z = 10 \times d$), where $\mathbf{e_z}$ denotes 
the unit normal vector in $z$-direction. 
At the outlet ($r = 5 \times d$), 
a fixed-value pressure is specified with 
the ``pressureInletOutletVelocity'' boundary condition
for flow velocity.
At the three-phase contact line, 
the ``dynamicAlphaContactAngle'' condition according to
\begin{equation}\label{dynamicContactAngle}
\theta = \theta_0 + (\theta_A - \theta_R) \tanh (u_w / u_{\theta})
\quad 
\end{equation}
is used  
with the static contact angle $\theta_0$, 
leading edge contact angle $\theta_A$, 
trailing edge contact angle $\theta_R$, and 
velocity scaling $u_{\theta}$ 
being specified.
The contact line moving velocity along the solid wall is 
denoted by $u_w$ in (\ref{dynamicContactAngle}), 
which becomes part of the solution when specifying
the contact angle condition related to $\nabla \alpha$ 
at the contact line that is implicitly allowed to move (or ``slip'')
in the local cell \citep[cf.][]{saha2009, linder2013}.
Without complete agreement on the boundary conditions
to be implemented at the moving contact line 
for modeling \citep{yarin2006},
the parameter values for ``dynamicAlphaContactAngle'' 
are selected somewhat arbitrarily in the present work 
only for demonstrating the possible fluid dynamics phenomena.

More often than not in 
Aerosol Jet$^{\circledR}$
operations,
the collimated mist stream is arranged to impinge perpendicularly 
onto the substrate.
With the mist stream wrapped in a thick gas sheath and
substrate typically located more than $10 \times$ 
the nozzle diameter away from nozzle exit, the individual ink droplets 
can be reasonably assumed to impact substrate perpendicularly 
with negligible deviations. 
Because the relevant droplet impact problem can be  
simplified to an axisymmetric configuration with a
simple rectangular domain as shown in figure 1, 
a mesh with wedge cells is generated with the {\em blockMesh} utility to
take advantage of axisymmetry for computations
of cases in the present study.
To ensure adequate resolution of the droplet free surface profile, 
the impaction region contains finite volume quadrilateral cells 
with side length less than
$0.01 \times d$ 
\citep[comparable to that used by][for VoF computations of 
drop impact problems]{toivakka2003, dinc2012}.
Initial position of the droplet center 
is set at $5 \times d$ 
with the droplet initial velocity 
set as the impinging gas jet velocity $U$, 
using the 
{\em funkySetFields} utility of ``swak4foam'' with 
OpenFOAM$^{\circledR}$.

For the nominal setting,
the surrounding gas (e.g., nitrogen---the typical mist carrier gas used in
Aerosol Jet$^{\circledR}$
process---under ambient temperature and pressure)
is assumed to have $\rho_g$ $= 1.2$ kg m$^{-3}$ 
and $\mu_g$ $= 1.8 \times 10^{-5}$ N s m$^{-2}$,
whereas the liquid droplet typically have $\rho_d$ $= 2 \times 10^3$ 
(but may vary 
between $1 \times$ and $4 \times$ $10^3$) kg m$^{-3}$
and $\mu_d$ in a range between $1 \times 10^{-3}$ and $1$
N s m$^{-2}$, representing the inks used in   
Aerosol Jet$^{\circledR}$
printing.
The surface tension of the droplet $\sigma$ is assumed to be constant
with a nominal value of $0.04$ 
(but may vary between $0.02$ and $0.08$) N m$^{-1}$.

As usual in fluid dynamics analysis,
nondimensional parameters can be conviniencely utilized.
If $\rho$ and $\mu$ are respectively measured in units of 
$\rho_d$ and $\mu_d$, length in units of $d$, 
velocity in units of $U$,
time in units of $d/U$,
and pressure in units of $\mu_d U/d$, 
three parameters would appear in 
the nondimensionalized (\ref{momentum}) 
such as the Reynolds number $Re$ $\equiv \rho_d U d/\mu_d$ 
in front of the first and second terms
on left side, 
the inverse capillary number $1/Ca$ $\equiv \sigma/(\mu_d U)$
in place of $\sigma$
and $\rho_d g d^2/(\mu_d U)$ $\equiv Bo/Ca$ 
in place of $g$ on right side, with 
$Bo$ denoting the Bond number $\rho_d g d^2/\sigma$.
Because the value of $Bo/Ca$ (as the ratio of 
the terminal velocity under gravity 
and impacting velocity $U$) even for 
a droplet of $d = 10^{-5}$ m, $\rho_d = 5 \times 10^3$ kg m$^{-3}$ 
and $\mu_d = 10^{-3}$
N s m$^{-2}$ 
at $U = 10$ m/s with $g = 9.81$ m s$^{-2}$ 
is $< 5 \times 10^{-3}$, 
the effect of gravity in  
Aerosol Jet$^{\circledR}$
ink droplet deposition 
(where $U$ is typically $> 20$ m/s) should be rather negligible.
Thus, only Reynolds number $Re$ ($\equiv \rho_d U d/\mu_d$) 
and capillary number $Ca$ ($\equiv \mu_d U/\sigma$) 
need to be specified as independent parameters
in computing the numerical results.

\section{Numerical results}\label{results}
Since the diameter of ink droplets rarely exceeds $5$ $\mu$m in
Aerosol Jet$^{\circledR}$
printing,
we start by examining cases with a droplet of $d = 5$ $\mu$m,
$\rho_d = 2 \times 10^3$ kg m$^{-3}$, 
$\mu_d = 10^{-3}$ 
N s m$^{-2}$ (or $1$ cp). 
As a reference, at $U = 100$ m/s (which represents the
high end of mist jet velocity in
Aerosol Jet$^{\circledR}$
printing) and $\sigma = 0.04$ N m$^{-1}$,
the value of $Re$ and $Ca$ are $1000$ and $2.5$, respectively.
When studying the droplet impact problem, many authors
often refer to the Weber number 
$We \equiv \rho_d U^2 d/\sigma$ $= Re \, Ca$ and 
the Ohnesorge number 
$Oh \equiv \mu_d/\sqrt{\rho_d \sigma d}$ 
$= \sqrt{Ca/Re}$ \citep{yarin2006}, 
which will also be used here as derived dimensionless
parameters in discussion. (Among $Re$, $Ca$, $We$, and $Oh$,
once two of them are specified as independent parameters 
the other two can then be determined from those specified two.) 
Corresponding to $Re = 1000$ and $Ca = 2.5$,
we have $We = 2500$ and $Oh = 0.05$ which represent 
cases of low viscosity ink drops of large sizes 
at high impact velocity 
relevant to  
Aerosol Jet$^{\circledR}$
printing.
In another extreme with an ink droplet of $d = 1$ $\mu$m,
$\rho_d = 1 \times 10^3$ kg m$^{-3}$, 
$\mu_d = 0.1$ 
N s m$^{-2}$ (or $100$ cp),
the values of $Re$ and $Ca$ for $\sigma = 0.08$ N m$^{-1}$ 
and impacting at $U = 20$ m/s become 
$0.2$ and $25$, yielding $We = 5$ and $Oh = 11.18$. 
Although dimensional parameter values are often referred to here, 
the results with plots are presented in terms of 
dimensionless parameters with length measured in units of $d$,
velocity in units of $U$, 
and time $t$ in units of $d/U$, for generality. 
The condition at contact line (\ref{dynamicContactAngle}) 
is specified as a static 
contact angle $\theta_0$ with  
$\theta_A = \theta_0 + 5^o$, $\theta_R = \theta_0 - 5^o$
and $u_{\theta} = 1$ m/s 
for {\em interFoam} computations.

With transient terms discretized using a first-order implicit Euler scheme,
the time step is controlled by setting the maximum Courant number to
$0.01$ (which is much finer than ``$< 0.5$'' as 
recommended by many authors to avoid significant spurious currents).
For postprocessing the numerical results, an open-source 
multi-platform data analysis package called {\em ParaView}
(available at www.paraview.org)
for scientific visualization is used in the present work.

\begin{figure}%[htb]
\centering
{\includegraphics[clip=true,scale=0.60,viewport=50 350 600 700]{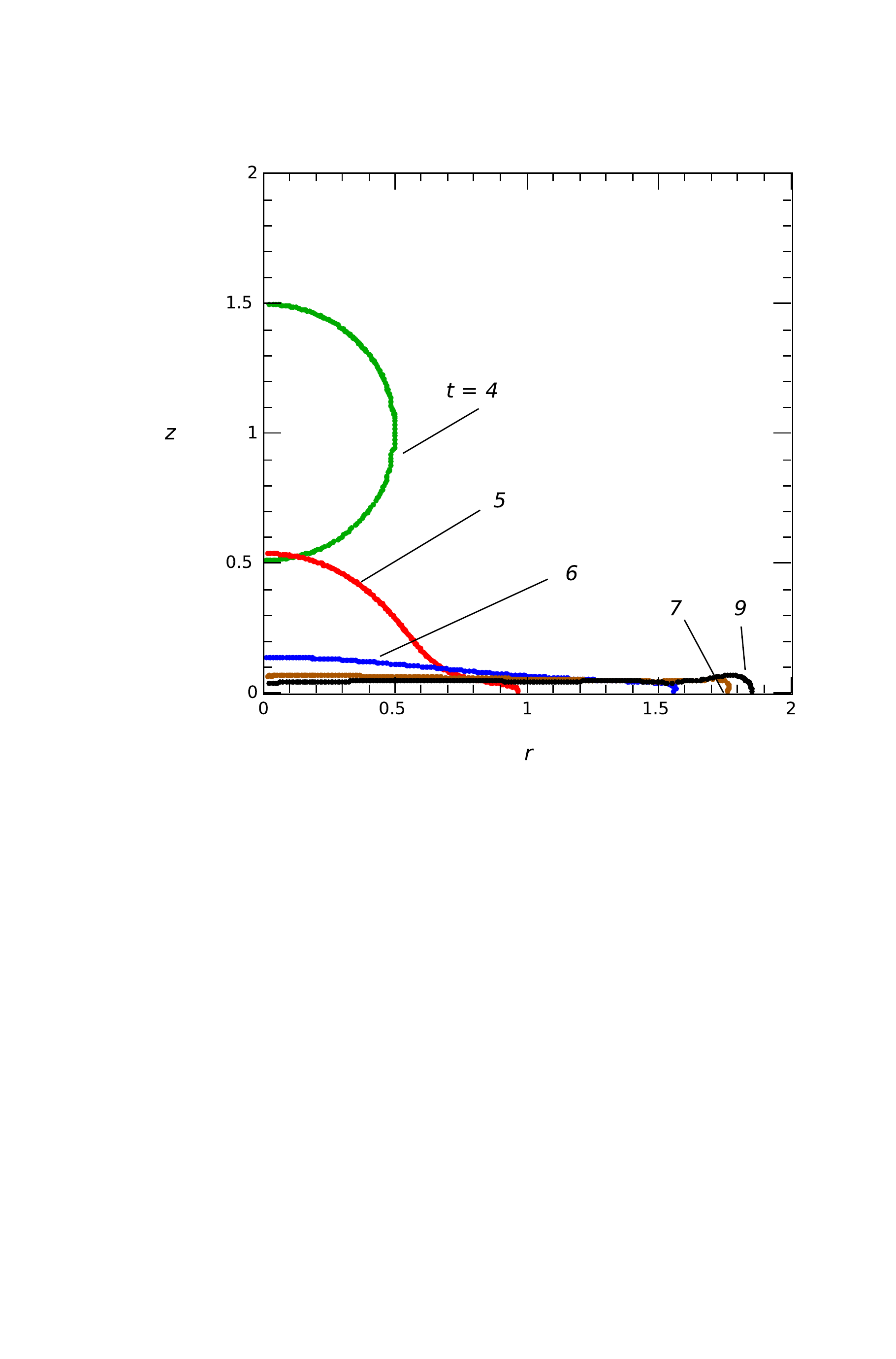}}
\caption{Spreading: shapes of a
droplet of $d = 5$ $\mu$m, $\rho_d = 2000$ kg m$^{-3}$,
and $\mu_d = 1$ cp with $\sigma = 0.04$ N m$^{-1}$, 
impacting solid surface
at $U = 100$ m/s ($Re = 1000$ and $Ca = 2.5$, 
or $We = 2500$ and $Oh = 0.05$) 
with contact angle $\theta_0 = 90^o$
for dimensionless time 
$t = 4$, $5$, $6$, $7$,
and $9$ in units of $d/U$ ($= 0.05$ $\mu$s),
from initial dimensionless position at $z = 5$ at $t = 0$.
The dimensionless coordinates $r$ and $z$ are 
labeled in units of
$d$.  The free surface profile data came from 
the output csv file of the {\em ParaView} contour for $\alpha = 0.5$.}
\label{fig:fig2}
\end{figure}

\subsection{Cases of $We = 2500$}
\label{We2500}
Droplets at large $We$ (e.g., $We = 2500$)
are expected to have 
relatively more significant dynamical deformations
and to exhibit more dramatic fluid dynamics phenomena.
Shown in figure 2 are the free surface profiles of 
a droplet impacting a solid surface and spreading 
as the contact radius increases with time
at $Re = 1000$ and $Ca = 2.5$,
for $\theta_0 = 90^o$ (with $\theta_A = 95^o$, $\theta_R = 85^o$,
and $u_{\theta} = 1$ m/s).
It appears that the droplet has little deformation before 
impacting the solid surface,
because the value of the Weber number based on gas flow 
$We_g$ $\equiv (\rho_g/\rho_d) \times Re \, Ca$ is only $1.5$,
as consistent with the findings of \cite{feng2010} that
noticeable deformations of liquid droplet moving 
in a gas medium are not expected for $We_g < 5$.
Soon after the droplet of $0.5$ radius 
(in units of $d$)
contacts the solid surface, 
it spreads to a maximum contact radius about $1.85$ 
at $t \approx 9$ (in units of 
$d/U = 0.05$ $\mu$s for $d = 5$ $\mu$m 
and $U = 100$ m/s).
It should be noted that the center of droplet is initially located at 
$z = 5$ at $t = 0$ 
moving at dimensionless velocity $1$ (in units of $U$ 
toward the solid surface (at $z = 0$);
therefore, 
the droplet bottom pole reaches the solid surface
around $t = 4.5$.
From $t = 4.5$ to $5$, the contact line moves from $r = 0$ to $r = 1$
(as indicated in figure 2)
with an estimated average speed of $\sim 2$ units of $U$.
Then, the speed of contact line motion is  
reduced to $\sim 0.6$ from $t = 5$ to $6$ 
and to $\sim 0.2$ from $t = 6$ to $7$ as the droplet becomes a lamella, 
and thereafter further down to $0$ at 
$t \approx 9$.
The time for spreading process, 
which is sometimes called the ``spreading time'' 
\cite[e.g., ][]{antonini2012},
is $\approx 4.5 \times d/U$ 
($= 0.225$ $\mu$s for $d = 5$ $\mu$m and $U = 100$ m/s). 

\begin{figure}%[htb]
\centering
{\includegraphics[clip=true,scale=0.60,viewport=50 350 600 700]{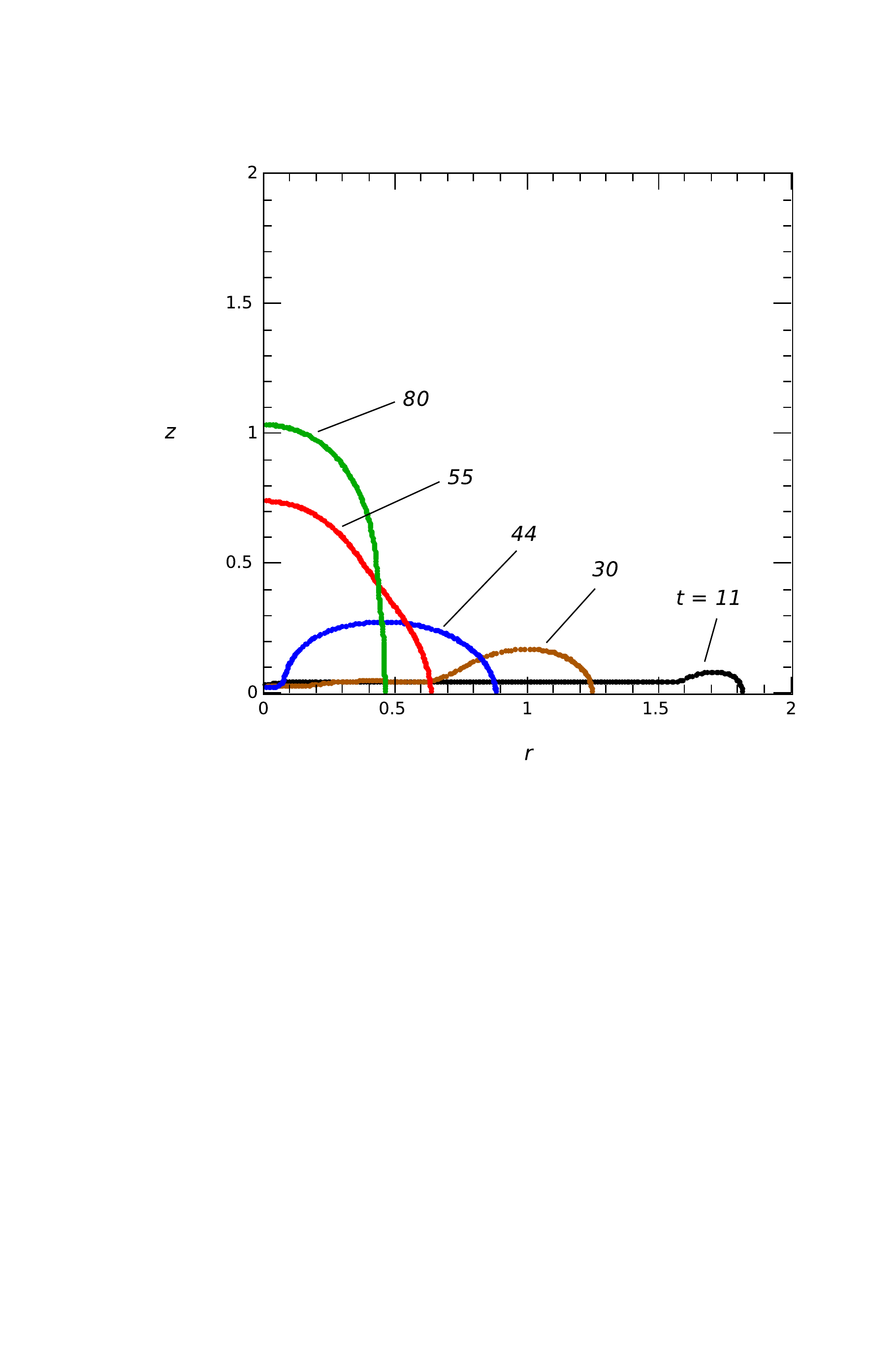}}
\caption{Receding: as in figure 2 but
for $t = 11$, $30$, $44$, 
$55$,
and $80$.}
\label{fig:fig3}
\end{figure}

\begin{figure}%[htb]
\centering
{\includegraphics[clip=true,scale=0.60,viewport=50 350 600 700]{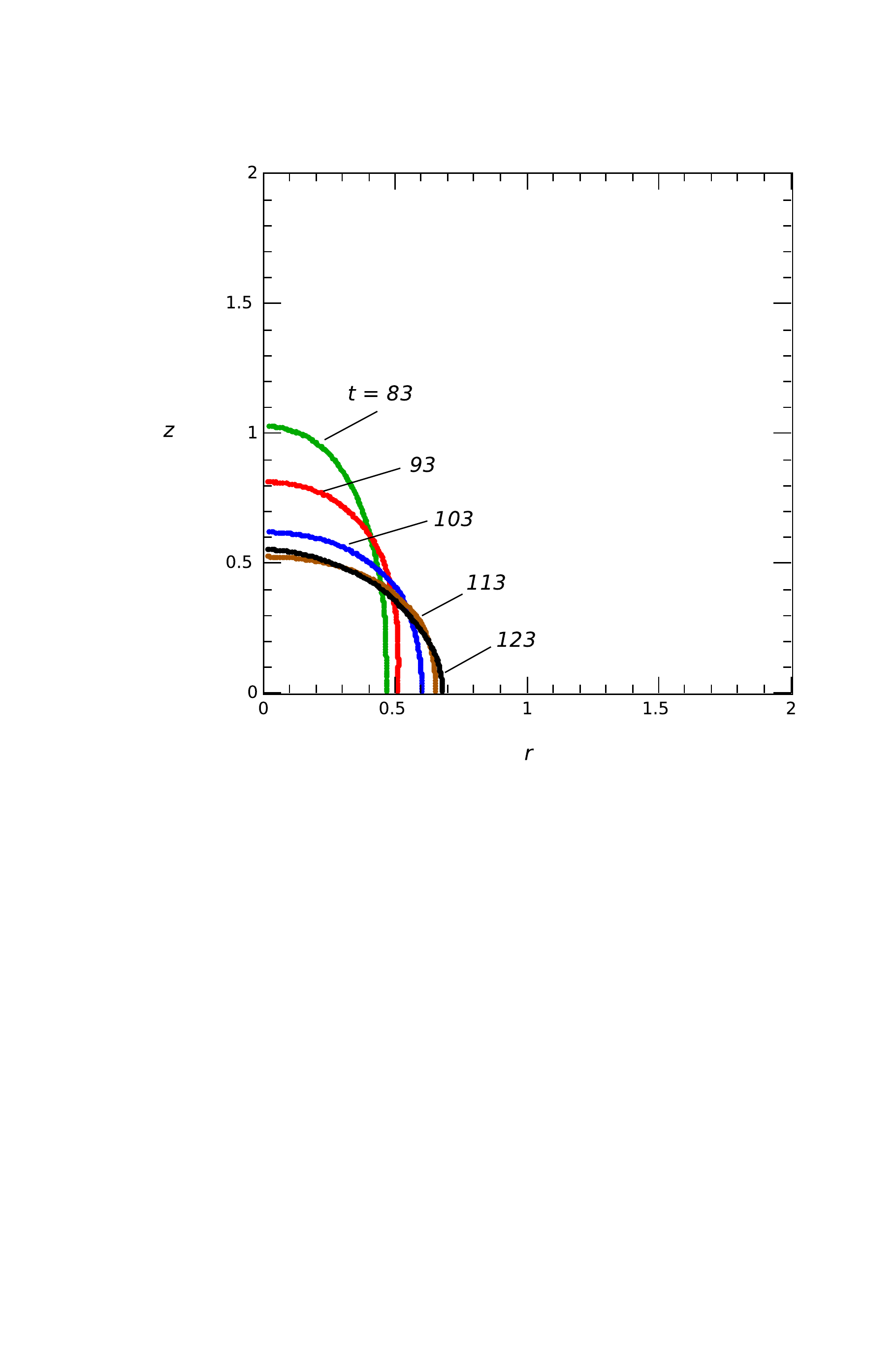}}
\caption{Oscillating: as in figure 2 but
for $t = 83$, $93$, $103$, 
$113$,
and $123$.}
\label{fig:fig4}
\end{figure}

After spreading to the maximum contact radius,
the lamella enters the receding (or relaxation) phase with
the contact radius shrinking with time from $t \approx 9$ 
until $t \approx 80$, 
as shown in figure 3
In the receding process, 
a growing bulged rim forms around the contact line 
as it moves toward the center 
at an average speed of $\sim 0.018$ units of $U$, 
much slower than that in the spreading process.
The inner edge of the bulged rim collapse at the center around 
$t = 44$ and thereafter the droplet center is pushed to move upward
quickly.
At $t = 55$, the contact line arrives the neighborhood of 
its equilibrium position $r \approx 0.63$.
Because of large $Re = 1000$, the contact radius 
continues to recede 
past its equilibrium position toward $r \approx 0.46$ at $t = 80$.
At the same time, the upper pole of free surface reaches its maximum 
height of $z \approx 1.03$.

What follows the receding of contact radius is
the sessile droplet oscillation up and down around its 
equilibrium hemispherical shape with a radius $\approx 0.63$,
due to relatively strong inertial effect at $Re = 1000$.
Even at $t = 103$ when the contact radius 
and upper pole are close to their 
equilibrium value ($0.63$), 
the free surface appears to still deviate noticeably from
its equilibrium hemispherical shape.
But the oscillation amplitude will decay quickly with time by 
viscous damping, which can clearly be seen in figure 5.

\begin{figure}%[htb]
\centering
{\includegraphics[clip=true,scale=0.60,viewport=50 340 600 720]{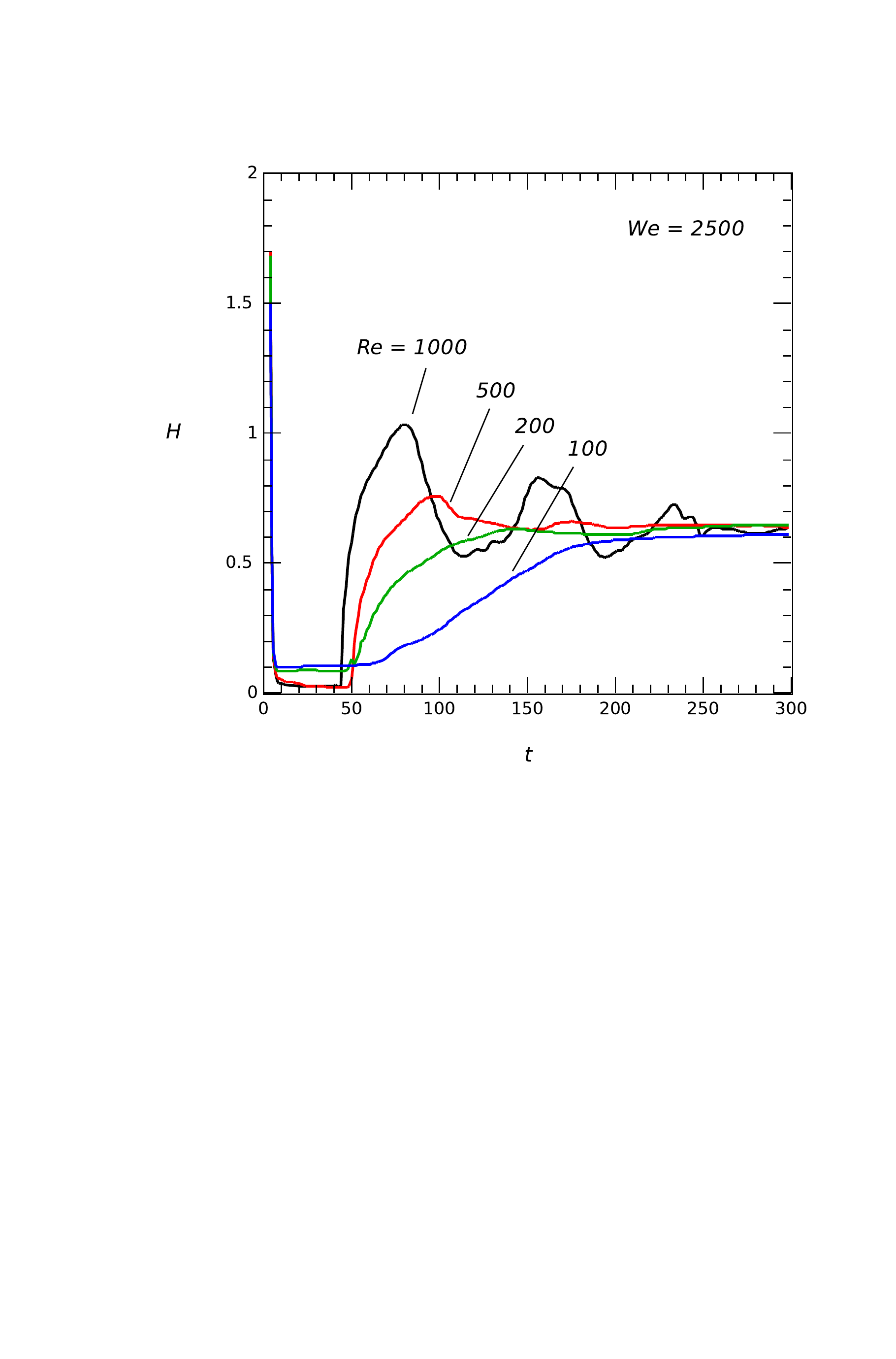}}
{\includegraphics[clip=true,scale=0.60,viewport=50 340 600 720]{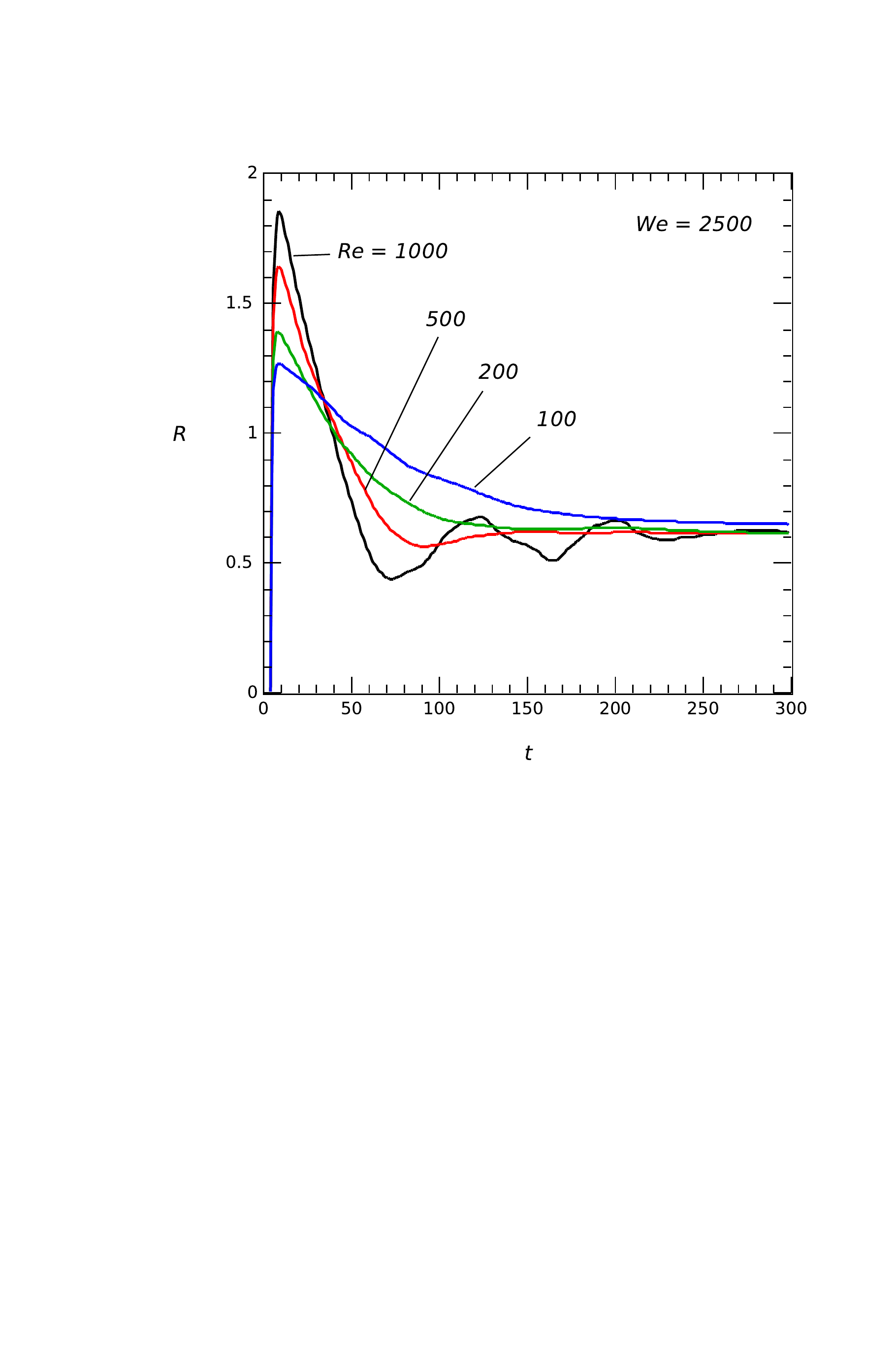}}
\caption{Plots of the center height $H$ 
($z$-value at $r = 0$ of free surface)
and contact radius $R$ ($r$-value at $z = 0$ of free surface) 
in units of $d$ versus time in units of $d/U$ ($= 0.05$ $\mu$s),
for droplets of $d = 5$ $\mu$m, $\rho_d = 2000$ kg m$^{-3}$,
$\sigma = 0.04$ N m$^{-1}$, 
with $\mu_d = 1$ cp ($Re = 1000$ and $Ca = 2.5$), 
$2$ cp ($Re = 500$ and $Ca = 5$), 
$5$ cp ($Re = 200$ and $Ca = 12.5$),
and $10$ cp ($Re = 100$ and $Ca = 25$),
impacting solid surface at $U = 100$ m/s
for contact angle $\theta_0 = 90^o$.}
\label{fig:fig5}
\end{figure}

The center height $H$ and contact radius $R$ (in units of $d$)
versus time are shown in figure 5 
for droplets of $d = 5$ $\mu$m, $\rho_d = 2000$ kg m$^{-3}$,
$\sigma = 0.04$ N m$^{-1}$ 
with $\mu_d = 1$ cp, $2$ cp, 
$5$, and $10$ cp,
impacting solid surface at $U = 100$ m/s.
For the case of $Re = 1000$ and $Ca = 2.5$ (corresponding to
that shown in figures 2--4),
the droplet exhibits significant oscillations after impaction.
Since the initial oscillation amplitude is quite large, 
the waveform does not appear to be simply sinusoidal
\citep[non-sinusoidal oscillations were also shown in 
experimental results of][]{ravi2010}. 
But the oscillation amplitude decays with time due to 
viscous damping. 
If the droplet viscosity is increased to $2$ cp 
(at $Re = 500$ and $Ca = 5$)
the amplitude of free surface oscillation diminishes rather quickly 
due to viscous damping after about 
one cycle.
Oscillations do not seem to occur for $\mu_d > 5$ cp 
(for $Oh > 0.25$)
indicating the deformed free surface after impaction 
creeps aperiodically to the equilibrium shape of a hemisphere.
After impaction, the center height $H$ appears to decrease with $t$
monotonically when $Re$ is large
reaching its minimium value right before the bulge rim collapse
at the center.
For example, the values of $H_{min}$ are 
$0.0266$ at $t = 36$ and 
$0.0231$ at $t = 41$ for $Re = 1000$ and $500$, respectively.
But with reducing the value of $Re$ (corresponding to increasing 
viscosity $\mu_d$) the value of $H_{min}$ tends to occur at 
smaller $t$ and then slowly increase.
For example, when $Re = 200$ and $100$
$H_{min} = 0.0858$ at $t = 9$ and 
$H_{min} = 0.0998$ at $t = 9$, respectively.

Immediately after the impaction, the contact radius $R$
(which is equivalent to one half of the so-called `spread factor') 
increases according to a square-root law,
such as $R \propto \sqrt{t^*}$ 
for $t^* \equiv t - t_0$ with $t_0$ denoting
the time for the impacting droplet to initiate contact to the substrate,
typically referred to as the kinematic phase 
when material points in the droplet mainly move in the 
$z-$direction rather than $r-$direction
\citep{rioboo2002}.  For the cases of $We = 2500$ in figure 5,
a curve fit of $R = 1.1 \sqrt{t^*}$ for $t_0 = 4.51$ 
appears to be quite accurate for $0 \le t^* \le 0.15$,
just as expected with the kinematic phase 
(usually considered for $t^* << 1$).
As a reference, some fittings of experimental data 
showed $R = 1.4 \sqrt{t^*}$ \citep{rioboo2002},
yet others had $R = 0.675 \sqrt{t^*}$ \citep[e.g.,][]{gupta2010}.
The droplet typically takes the shape 
of a `cut sphere' during the kinematic phase,
similar to that at $t = 5$ in figure 2,
until a  lamella---radially expanding film
bounded by a rim---forms in the spreading phase 
(like the profiles shown in figure 2 for $t > 6$). 
All curves of $R$ versus $t$ in figure 5 exhibit a common feature 
with a quick increase of $R$ immediately after the impact 
at $t \sim 4.5$ until $t \sim 9$ where $R$ reaches a peak value,
and then $R$ decreases with $t$ at a much slower speed.
The spreading phase ends when the spreading velocity 
approaches zero, which usually corresponds to 
$R$ arriving at its peak value.
Following the spreading phase the lamella may begin to 
recede, which is sometimes called the relaxation phase 
because the receding contact line now is moving at a 
relatively much lower speed (as shown in figure 3 and $R(t)$ in
figure 5).
After the relaxation phase, 
the impact kinetic energy is almost dissipated by 
the viscous effect, and 
the droplet will go through a 
slow lengthy `wetting equilibrium' phase 
\citep[e.g., $R(t) \propto t^{1/10}$ according to][]{tanner1979}
toward the capillary equilibrium determined by the static contact angle.

Among many variables involved in droplet impact dynamics, 
the maximum spread factor, characterizing the maximum 
value of contact diameter normalized with the diameter
of the undeformed droplet before impaction, 
has often been considered in the literature 
for describing the impaction dynamics as well as for comparing
results. 
The maximum spread factor can be brought to bear on 
various practical applications such as 
inkjet printing, spray coating, pesticide application, etc.
where the actual droplet coverage area 
directly corresponds to maximum spreading due to 
rapid solidification at contact line, 
liquid absorption into porous substrate,
contact line pinning on a textured surface, and so on so forth.
According to an empirical correlation by 
\cite{scheller1995},
the maximum spread factor $\xi \equiv 2 R_{max}$ 
may be expressed as \citep[cf. ][]{yarin2006}
\begin{equation}\label{spread-factor}
\xi = 0.61 \times (Re^2 \, Oh)^{0.166} 
\quad .
\end{equation}
Another semiempirical relation was proposed by \cite{roisman2009}
which, out of numerous possibilities, 
is presented here in a modified form as
\begin{equation}\label{spread-factor2}
\xi = Re^{1/5} - 0.35 Re^{2/5} / \sqrt{We} 
\quad ,
\end{equation}
with the original factors $0.87$ and $0.40$ replaced here 
by $1.0$ and $0.35$ for
a better match to the values of present computational results. 

A comparison of 
(\ref{spread-factor}) and 
(\ref{spread-factor2})
with the present results
is given in table~\ref{t1} (for a constant $We = 2500$),
which shows remarkably good agreements.
As reasonably accurate as they may seem though,
neither (\ref{spread-factor}) nor (\ref{spread-factor2}) 
explicitly accounts for 
the contact angle effect, 
which was somehow justified by experiments 
\citep[cf.][]{scheller1995}.
According to \cite{rioboo2002}, 
immediately after the droplet touches the substrate
(and thoughout most part of the spreading phase)
the contact line motion is controlled by the dominant 
kinetic energy, irrespective of the physical properties of
the liquid and solid such as the contact angle.
Thus, contact angle 
may not be expected to have significant effect on 
the dynamics of spreading following the droplet impact 
and the value of 
maximum contact radius $R$ at the end of spreading.

\begin{table}
\caption{\label{t1}Comparison of the present computed values of 
the maximum spread factor $\xi$
with that predicted by (\ref{spread-factor}) at $We = 2500$, 
for droplets of $d = 5$ $\mu$m, $\rho_d = 2000$ kg m$^{-3}$,
$\sigma = 0.04$ N m$^{-1}$
with various values of viscosity $\mu_d$, 
when impacting solid surface at $U = 100$ m/s
for contact angle $\theta_0 = 90^o$ (with $\theta_A = 95^o$
and $\theta_R = 85^o$).}
%\begin{center}
\begin{indented}
\item[]\begin{tabular*}{0.75\textwidth}{@{\extracolsep{\fill}} c c c c c c c}
%\hline
\br
\\
$\mu_d$ (cp) & $Re$ & $Ca$ & $Oh$ & $\xi$ & Eq. (\ref{spread-factor}) & Eq. (\ref{spread-factor2}) \\
%\hline
\mr
1 & 1000 & 2.5 & 0.05 & 3.696 & 3.676 & 3.870 \\
2 & 500  &  5 & 0.1 & 3.281 & 3.276 & 3.382 \\
5 & 200 & 12.5 & 0.25 & 2.796 & 2.814 & 2.827 \\
10 & 100 & 25 & 0.5 & 2.462 & 2.508 & 2.468 \\
100 & 10 & 250 & 5 & 1.714 & 1.711 & 1.567 \\
%\hline
\br
\end{tabular*}
%\end{center}
\end{indented}
\end{table}

To test the validity of 
(\ref{spread-factor}) and
(\ref{spread-factor2}) 
for contact angles
other than $\theta_0 = 90^o$, computations of 
cases for $\theta_0 = 45^o$ (with $\theta_A = 50^o$
and $\theta_R = 40^o$) while other parameters remain 
unchanged from those in table 1) are also performed.
The results show that $\xi = 3.827$ 
for $\mu_d = 1$ cp ($Re = 1000$), 
$3.396$ for $2$ cp ($Re = 500$), 
$2.849$ for $5$ cp ($Re = 200$), 
$2.484$ for $10$ cp ($Re = 100$), 
and $1.724$ for $100$ cp ($Re = 10$), with a spreading time 
$\approx 6.5 \times d/U$ ($=0.325$ $\mu$s) 
which is about $0.1$ $\mu$s longer than the case of $\theta_0 = 90^o$
for droplets of $d = 5$ $\mu$m at $U = 100$ m/s,
Results for
$\theta_0 = 135^o$ (with $\theta_A = 140^o$
and $\theta_R = 130^o$) 
show that $\xi = 3.624$ 
for $\mu_d = 1$ cp ($Re = 1000$), 
$3.231$ for $2$ cp ($Re = 500$), 
$2.782$ for $5$ cp ($Re = 200$), 
$2.448$ for $10$ cp ($Re = 100$), 
and $1.705$ for $100$ cp ($Re = 10$), with a spreading time 
$\approx 3.5 \times d/U$ ($=0.175$ $\mu$s).
Hence, the computed values of $\xi$ (at $We = 2500$)
are indeed insensitive to the contact angle variations,
as consistent with the experimental findings of 
\cite{scheller1995}.

However, the contact angle may 
drastically influence the dynamics of free surface deformation
after the completion of 
spreading phase \citep[as described by][]{rioboo2002}.  
Figure 6 shows that a droplet with 
contact angle $\theta_0 = 45^o$ 
(at $Re = 1000$ and $Ca = 2.5$)
recedes very slowly 
in contrast to the case of $\theta_0 = 90^o$
with considerable oscillations after receding, 
while the droplet with $\theta_0 = 135^o$ recedes rapidly with 
great momentum such that it bounces off the solid surface 
(around $t = 51$).

\begin{figure}%[htb]
\centering
{\includegraphics[clip=true,scale=0.56,viewport=50 340 600 720]{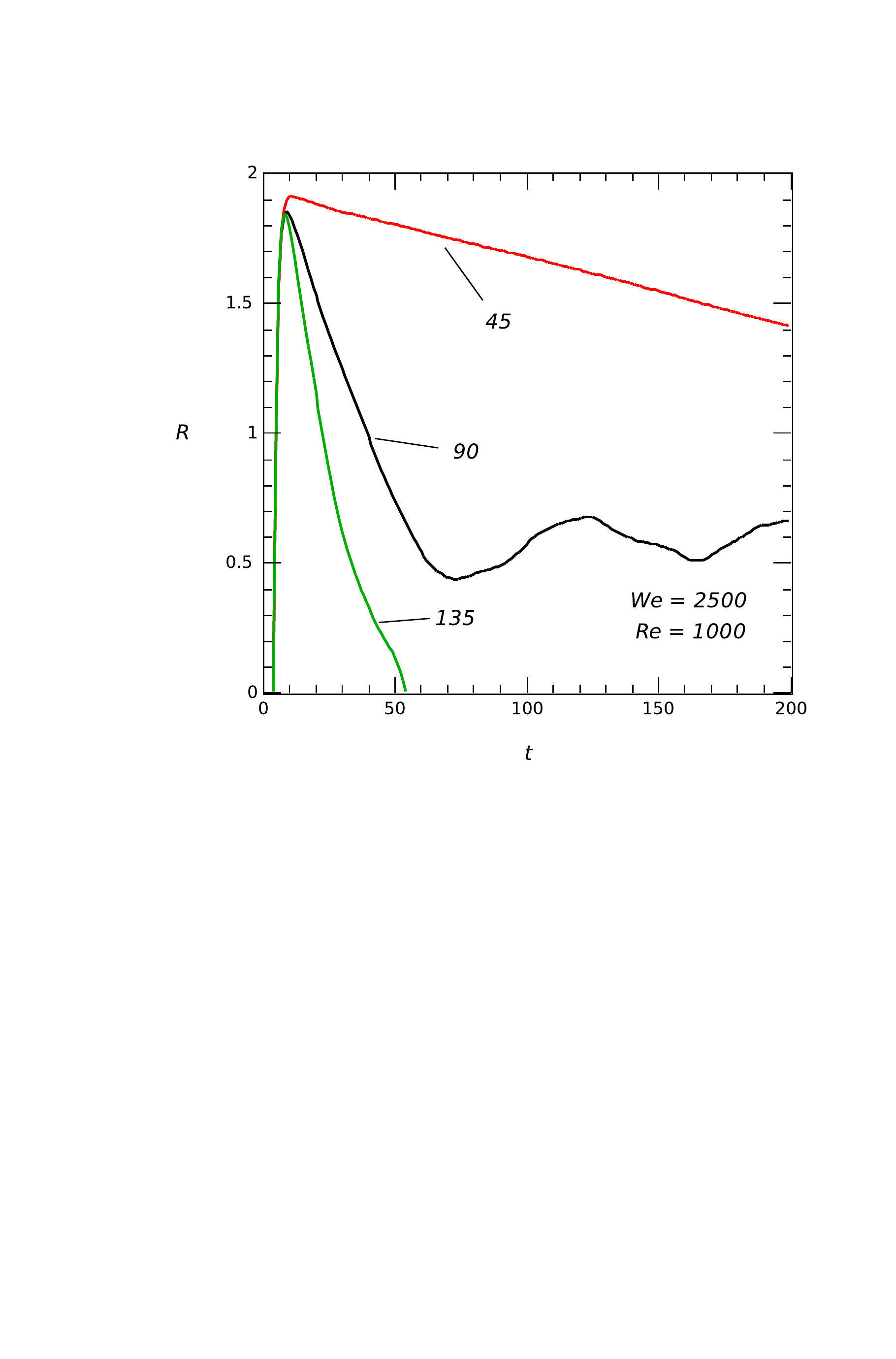}}
\caption{Plot of 
contact radius $R$, as in figure 5,
with $\mu_d = 1$ cp ($Re = 1000$ and $Ca = 2.5$), 
but for contact angle $\theta_0 = 45^o$
$90^o$, and $135^o$, as labeled.}
\label{fig:fig6}
\end{figure}

Physically, bouncing can occur when kinetic energy
of impact remains sufficiently large in the receding phase
if the viscous dissipation effect is relatively weak
such that the shrinking lamella contact line virtually 
disappears near the impact point \citep{yarin2006}.
Figure 7 shows the snapshots of such free surface shape evolution 
from spreading to receding-bouncing, with streamlines being
also displayed to illustrate external gas flow field 
interaction with the free surface deformation at 
different stages. Clearly, 
the liquid droplet impact dynamics can influence
the external gas flow significantly.
In view of the gas flow streamline configuration,
the droplet, having about the same velocity 
as the surrounding gas at $t = 0$,
moves ahead the decelerating gas due to its inertia
as it approaches the substrate
at $t = 3$. 
The fast moving liquid phase tends to drag nearby gas at 
an increased velocity 
during the droplet spreading phase, e.g., at $t = 5$.
Toward the end of droplet spreading phase, e.g., at $t = 8$,
as the liquid phase motion decreases
the gas phase around free surface recovers its natural 
impinging jet type of flow.
During the relatively slow receding process, e.g., at $t = 15$
a somewhat stagnant 
zone in the gas phase develops near the free surface.
As the droplet leaving the substrate during bouncing,
a low velocity wake appear behind it, e.g, at $t = 51$ and $80$.
The bouncing droplet is expected to experience a 
gas flow resistance that tends to 
push it back toward the substrate.
Eventually, the bouncing droplet will come back
to reattach to the substrate due to the external gas flow of 
an impinging jet.

\begin{figure}%[htb]
\centering
{\includegraphics[clip=true,scale=0.66,viewport=163 510 383 740]{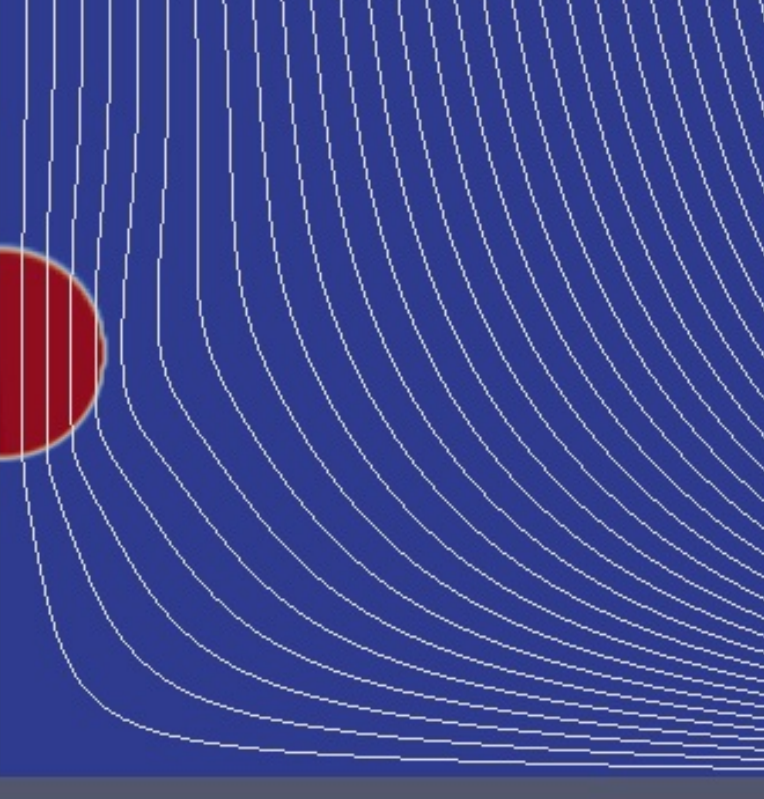}}
{\includegraphics[clip=true,scale=0.66,viewport=163 510 383 740]{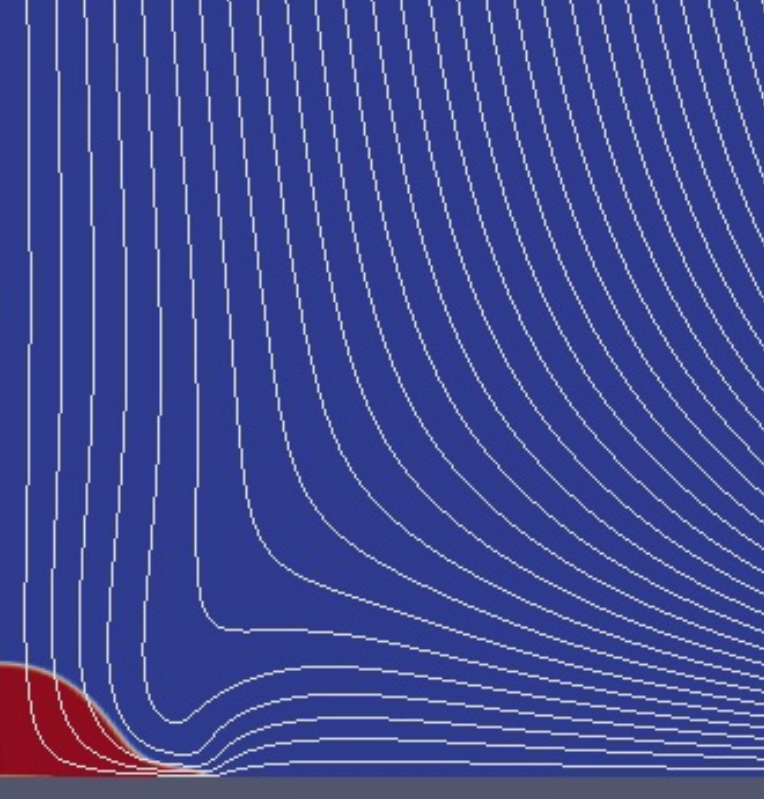}}
{\includegraphics[clip=true,scale=0.66,viewport=163 510 383 740]{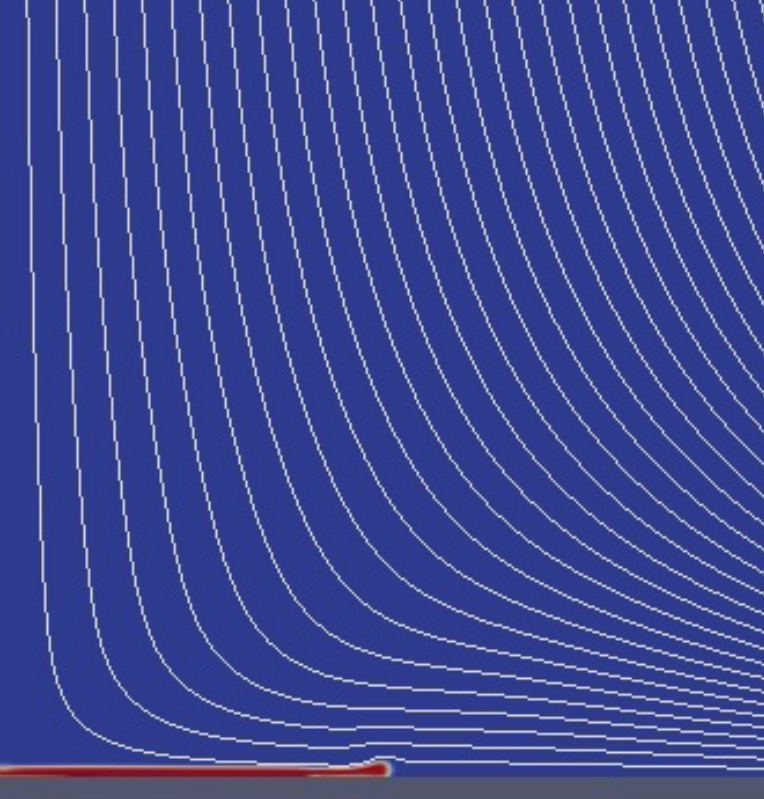}}
\vspace*{2mm}
\vspace*{2mm}
{\includegraphics[clip=true,scale=0.66,viewport=163 510 383 740]{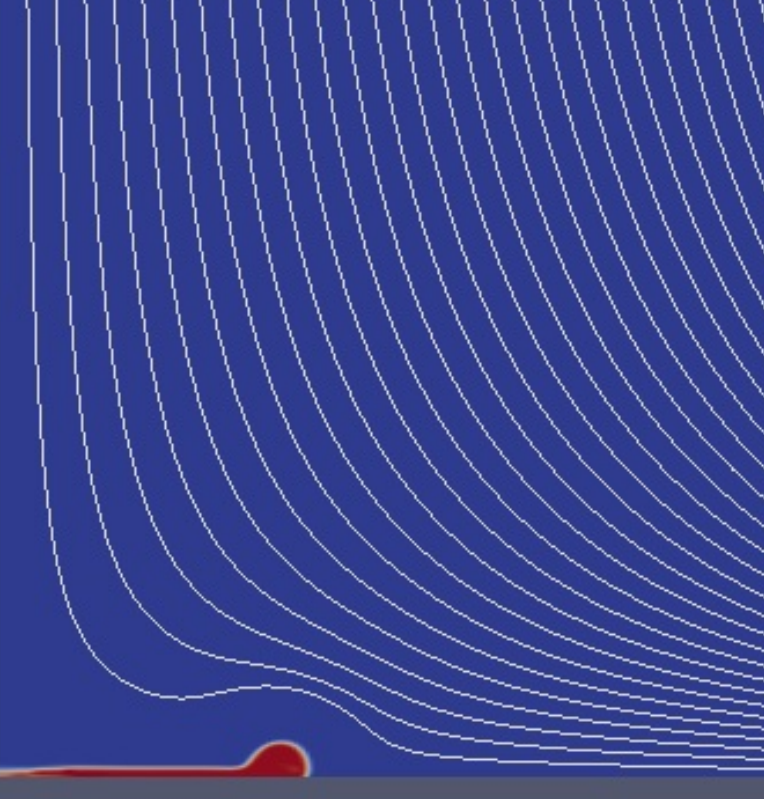}}
{\includegraphics[clip=true,scale=0.66,viewport=163 510 383 740]{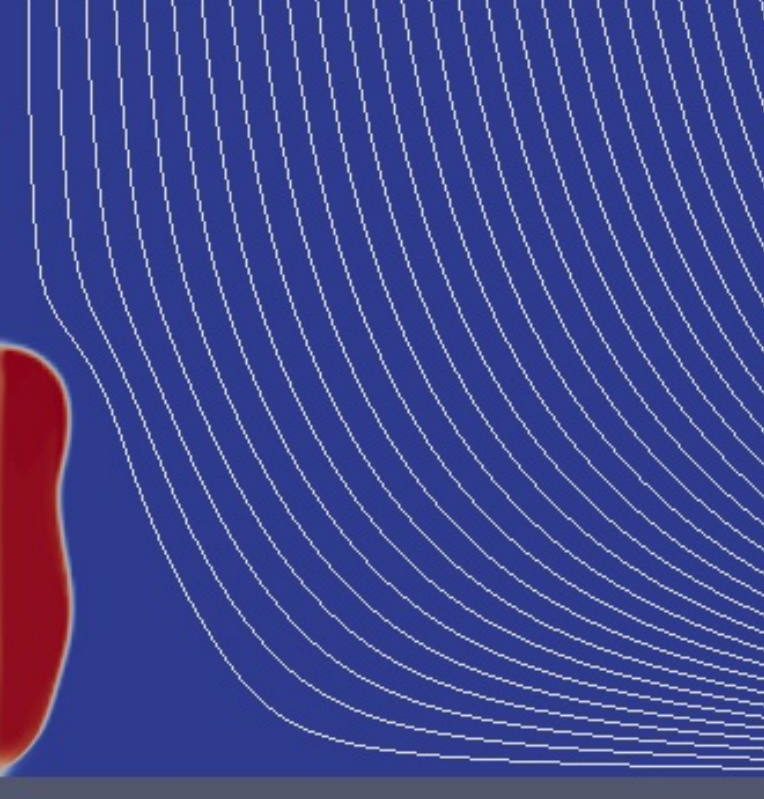}}
{\includegraphics[clip=true,scale=0.66,viewport=163 510 383 740]{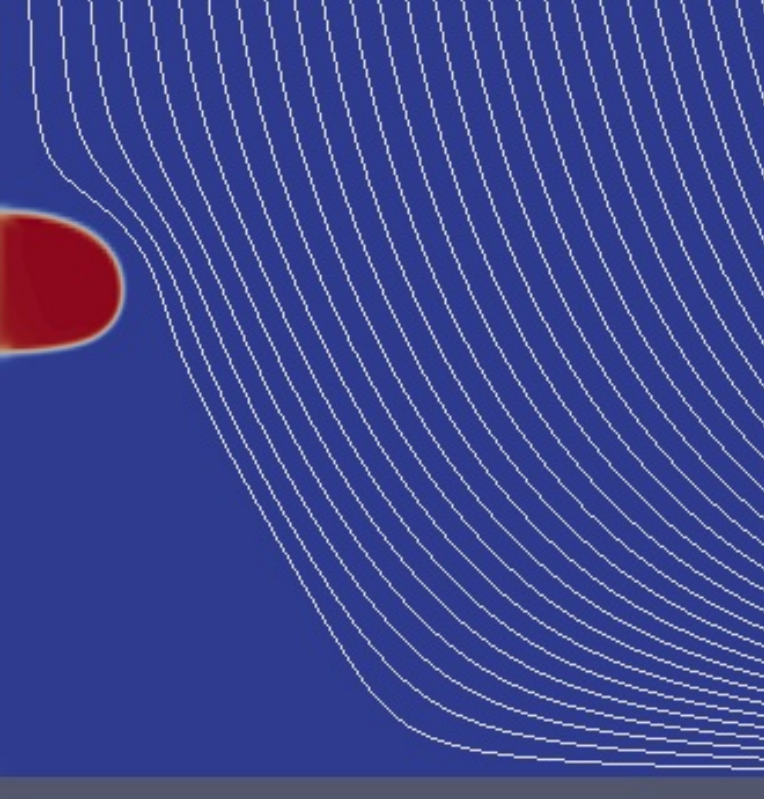}}
\caption{Streamlines associated with a droplet impaction 
at $Re = 1000$ and $Ca = 2.5$ 
($We = 2500$ and $Oh = 0.05$, as in figure 2), 
but for contact angle
$\theta_0 = 135^o$ ($\theta_A = 140^o$, $\theta_R = 130^o$)
for spreading at $t = 3$, $5$, $8$ in the upper row, 
and receding to bouncing at $t = 15$, $51$, and $80$ in the lower row.} 
\label{fig:fig7}
\end{figure}

An examination of the effect of liquid viscosity $\mu_d$ 
for a droplet with contact angle $\theta_0 = 135^o$
indicates that the dynamics after spreading 
is also controlled by the value of Reynolds number $Re$.
Figure 8 shows that receding momentum decreases with 
increasing $\mu_d$ (namely, reducing $Re$),
and bouncing would not occur when $Re = 250$
(for $R$ never reaches zero before recoving to a spreading phase again).
At $Re = 333$ the droplet detaches from the substrate around $t = 93$,
but reattaches to the substrate around $t = 143$ for lack of bouncing
momentum.
Reducing $Re$ tends to increase
the time from the droplet impact to its detaching from substrate, 
if bouncing occurs.
For example, a droplet with $Re = 1000$ 
impacts the substrate at $t = 4.5$ and detaches from the substrate 
at $t \approx 53$,
with $Re = 500$ at $t \approx 63$,
and with $Re = 333$ at $t \approx 96$.   
As expected, liquid viscosity has an effect of retarding 
the free surface flow during receding-bouncing.

\begin{figure}%[htb]
\centering
{\includegraphics[clip=true,scale=0.56,viewport=50 340 600 720]{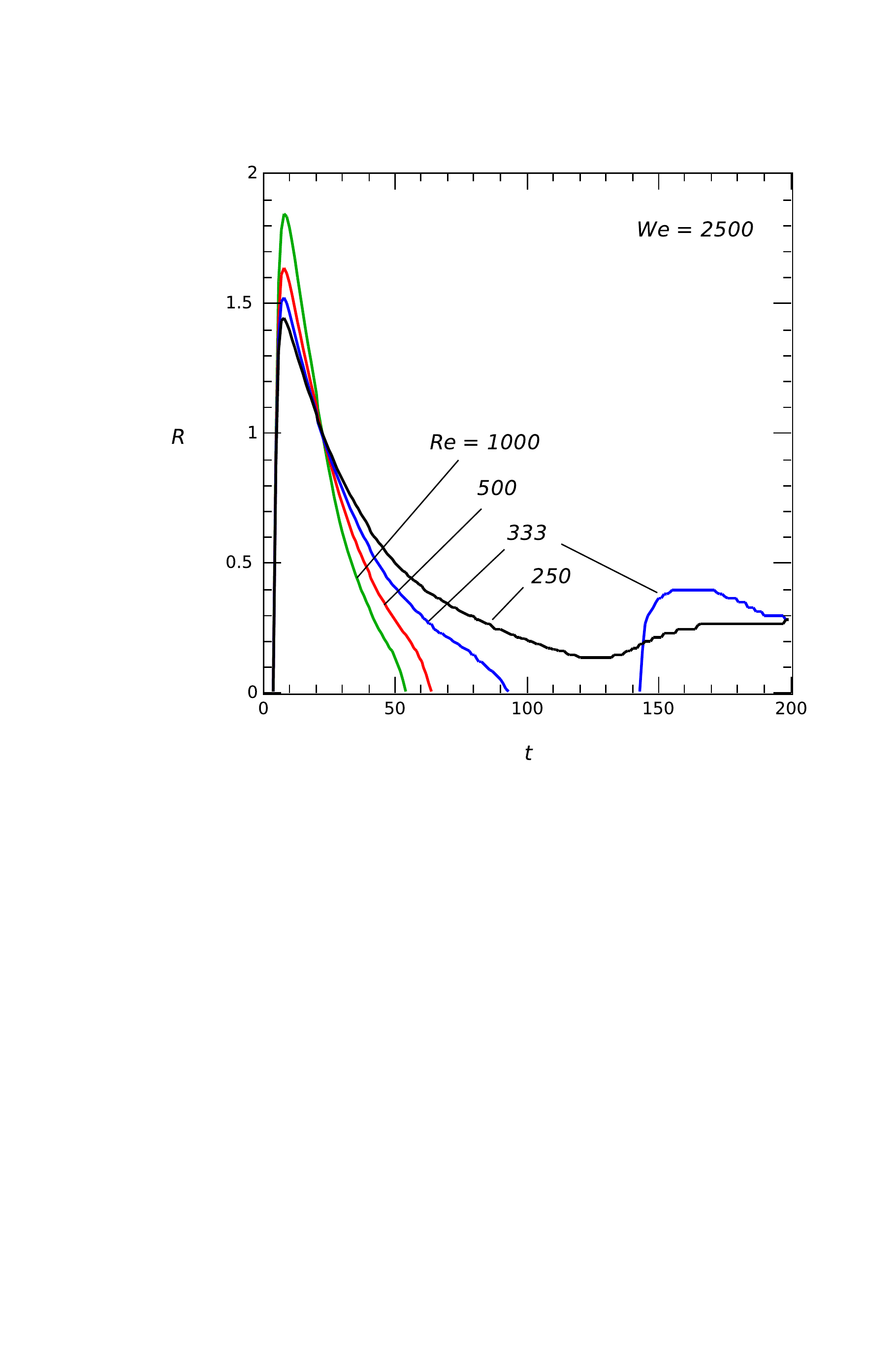}}
\caption{Plot of 
contact radius $R$, as in figure 6,
with $\mu_d = 1$ cp ($Re = 1000$ and $Ca = 2.5$), 
$\mu_d = 2$ cp ($Re = 500$ and $Ca = 5.0$),
$\mu_d = 3$ cp ($Re = 333$ and $Ca = 7.5$), and
$\mu_d = 4$ cp ($Re = 250$ and $Ca = 10$), 
for contact angle $\theta_0 = 135^o$.} 
\label{fig:fig8}
\end{figure}

\begin{figure}%[htb]
\centering
{\includegraphics[clip=true,scale=0.60,viewport=50 350 600 700]{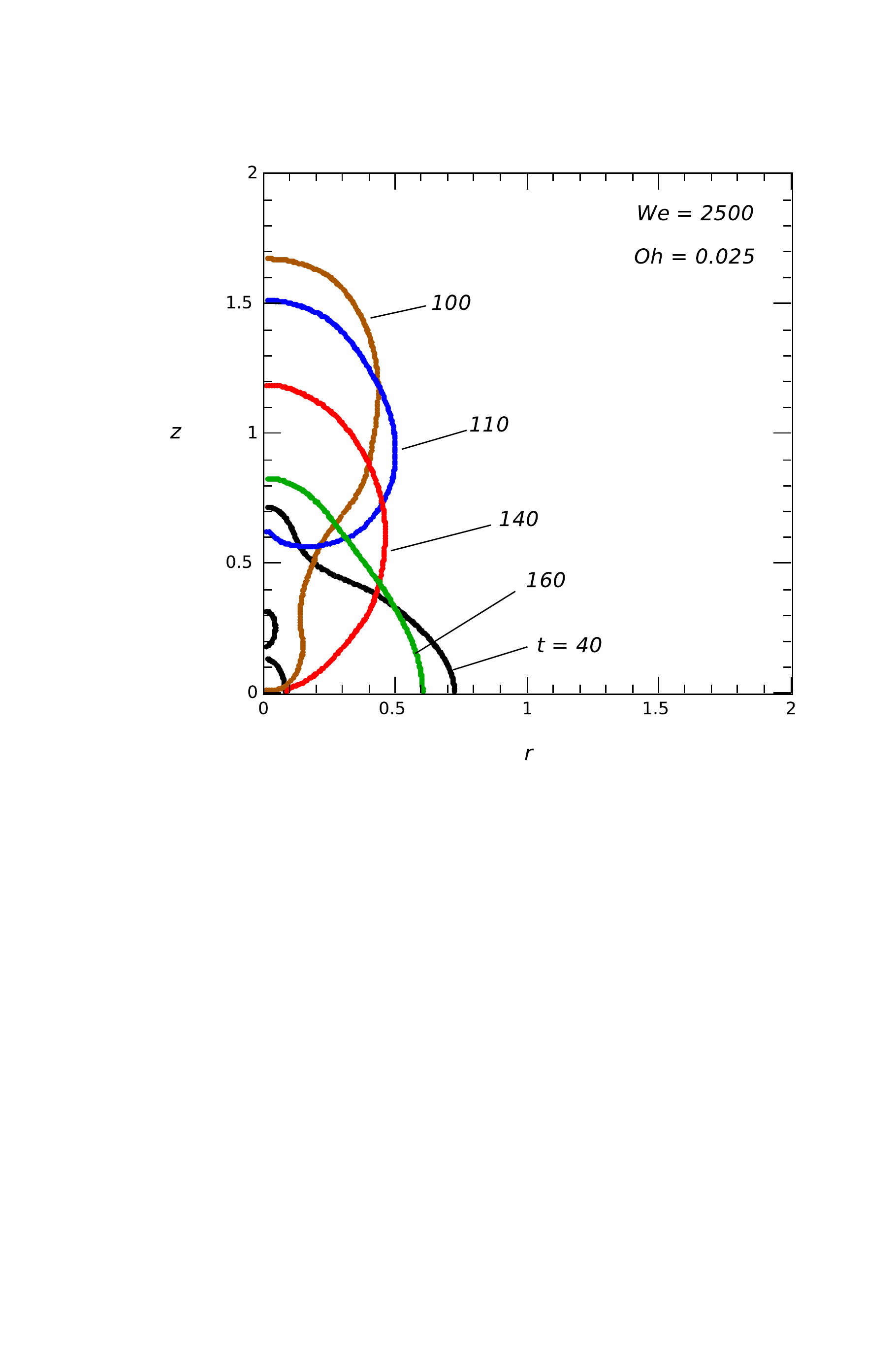}}
\caption{Detaching and reattaching: as in figure 2 but
for $\rho_d = 4000$ kg m$^{-3}$ and surface tension
$\sigma = 0.08$ N m$^{-1}$
($Re = 2000$ and $Ca = 1.25$, or $We = 2500$ and $Oh = 0.025$)
with $\theta_0 = 90^o$
at $t = 40$, $100$, $110$, 
$140$,
and $160$.}
\label{fig:fig9}
\end{figure}

If a droplet of $d = 5$ $\mu$m with viscosity $\mu_d = 1$ cp has 
density $\rho_d = 4000$ kg m$^{-3}$ and surface tension
$\sigma = 0.08$ N m$^{-1}$,
the values of $Re$ and $Ca$ for $U = 100$ m/s become
$2000$ and $1.25$, which lead to $We = 2500$ and 
$Oh = 0.025$.
The computed maximum spread factor is $\xi = 4.177$,
while (\ref{spread-factor})
and (\ref{spread-factor2}) predict $4.124$ and $4.4267$, respectively.
This may be considered as an extreme case for
relatively weak droplet viscosity effect 
compared to the inertial and surface tension effects
directly relevant to
Aerosol Jet$^{\circledR}$
droplet deposition. 
Because of the reduced viscous effect, 
enhanced free surface deformations can be observed.
Figure 9 shows a phenomenon of detaching and reattaching, 
after a droplet impact for $Re = 2000$ and $Ca = 1.25$
(or $We = 2500$ and $Oh = 0.025$) with $\theta_0 = 90^o$.
Entrapped bubbles can be seen to form
during the receding phase, as indicated in the free surface 
profile at $t = 40$.
The center height $H$ reaches its peak value $1.779$ at 
$t = 83$. 
The contact radius $R$ shrinks to zero at $t = 106$ 
corresponding to the time for complete detachment of the droplet
from substrate,
when the free surface pinches off 
at the end of a tail formed at the droplet bottom (cf. the 
free surface profile at $t = 100$). 
The tip of such tail moves rapidly upward into 
the bulk of the droplet, due to the action of surface tension,
leaving a deep dimple on the droplet bottom at $t = 110$.  
While moving downward and oscillating with a considerable amplitude,
the bottom of the detached droplet reattaches the 
substrate at $t = 140$.
Thereafter, the attached droplet exhibits 
significant oscillatory motions with even larger amplitudes 
than that shown in figure 4, as a consequence of relatively
stronger effects of fluid inertia and surface tension.

\subsection{Cases of $We = 100$}
\label{We100}
If the droplet of $d = 5$ $\mu$m, $\rho_d = 2000$ kg m$^{-3}$,
$\mu_d = 1$ cp, 
and $\sigma = 0.04$ N m$^{-1}$
has an impact velocity of $20$ m/s,
the values of $Re$ and $Ca$ become $200$ and $0.5$ such that
$We = 100$ and $Oh = 0.05$.
The spreading time for $We = 100$ is around $t = 2.0$,
corresponding to a dimensional time of $\approx 2.0 \times d/U$ 
($= 0.5$ $\mu$s for $d = 5$ $\mu$m and $U = 20$ m/s),
which seems to be consistent with the experimental findings of 
drop impact scaling time
$t \times d/U \propto We^{-0.25}$ by
\cite{antonini2012},

The effect of increasing liquid viscosity $\mu_d$ on dynamics of 
a droplet of $d = 5$ $\mu$m,
$\rho_d = 2000$ kg m$^{-3}$,
and $\sigma = 0.04$ N m$^{-1}$
with an impact velocity of $20$ m/s
is shown in figure 10 for 
the center height $H$ and contact radisu $R$ versus time.
It is interesting to note that the curves in figure 10
are quite similar to those in figure 5 
corresponding to the same values of $Oh$,
despite more than an order of magnitude reduction of 
$We$.
Because
there is a factor of $5$ difference in the reference time scale
$d/U$ (due to a factor of $5$ reduction of $U$),
the normalized time interval $(0, 60)$ for $t$ in figure 10 
has the same dimensional time interval (0, 15$\mu$s) 
as $(0, 300)$ for $t$ in figure 5. 
This is expected in view of the fact that the impact velocity $U$
only provides the initial free surface deformation that sets 
the droplet into free oscillatory motion, the characteristics of which is 
usually determined by fluid density $\rho_d$, droplet size $d$, 
and surface tension $\sigma$ \citep{landau1959}.
Noteworthy here is that the Ohnesorge number $Oh$ 
($\equiv \mu_d/\sqrt{\rho_d \sigma d}$)
is independent of $U$, unlike $Re$, $Ca$, and $We$.
The fluid viscosity contained in $Oh$ is responsible for the decay of
oscillation amplitude, whereas $\rho_d$, $d$, and $\sigma$ are the 
key ingredients for capillary driven oscillations.
Similar to the case of $We = 2500$, 
the oscillatory motion seems to also diminish for $Oh > 0.25$ 
at $We = 100$.

Similar to that shown in figure 5, 
the center height $H$ decrease with $t$ after impaction.
But the lamella at $We = 100$ is thicker (with larger $H_{min}$) than 
that corresponding to the same value of $Oh$ at $We = 2500$.
For example, the values of $H_{min}$ are reached as
$0.0353$ at $t = 9.6$ for $Re = 200$, 
$0.1097$ at $t = 7.8$ for $Re = 100$, 
$0.1940$ at $t = 6.8$ for $Re = 40$, and 
$0.2881$ at $t = 6.2$ for $Re = 20$ and $500$, respectively.
While the oscillatory characteristics following the spreading phase 
appear independent of the impact velocity $U$ (and the value of $We$),
the thickness of the spreading lamella as well as the maximum
spread factor are strongly influenced by the value of $We$.

\begin{figure}%[htb]
\centering
{\includegraphics[clip=true,scale=0.60,viewport=50 340 600 750]{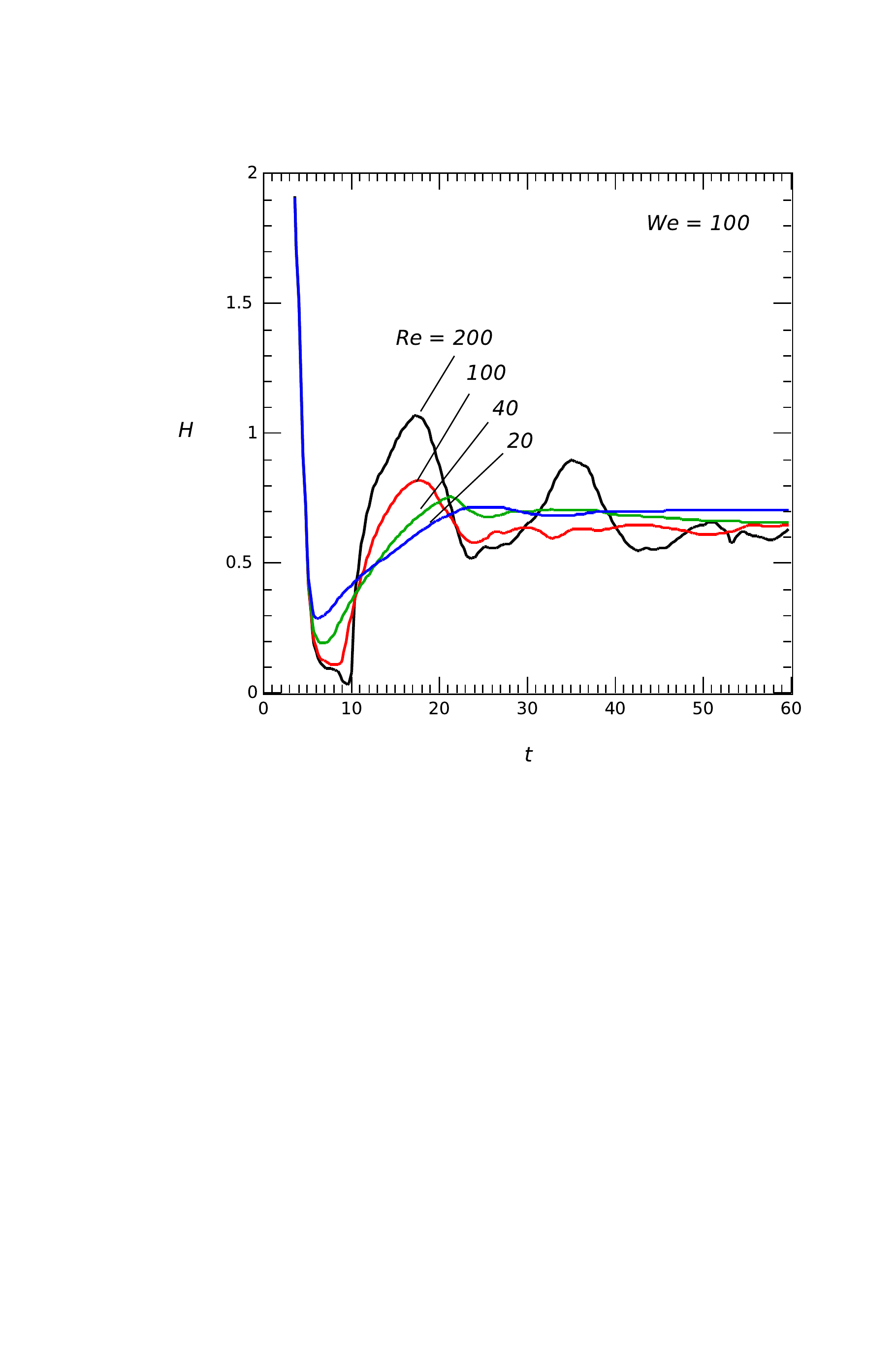}}
{\includegraphics[clip=true,scale=0.60,viewport=50 340 600 750]{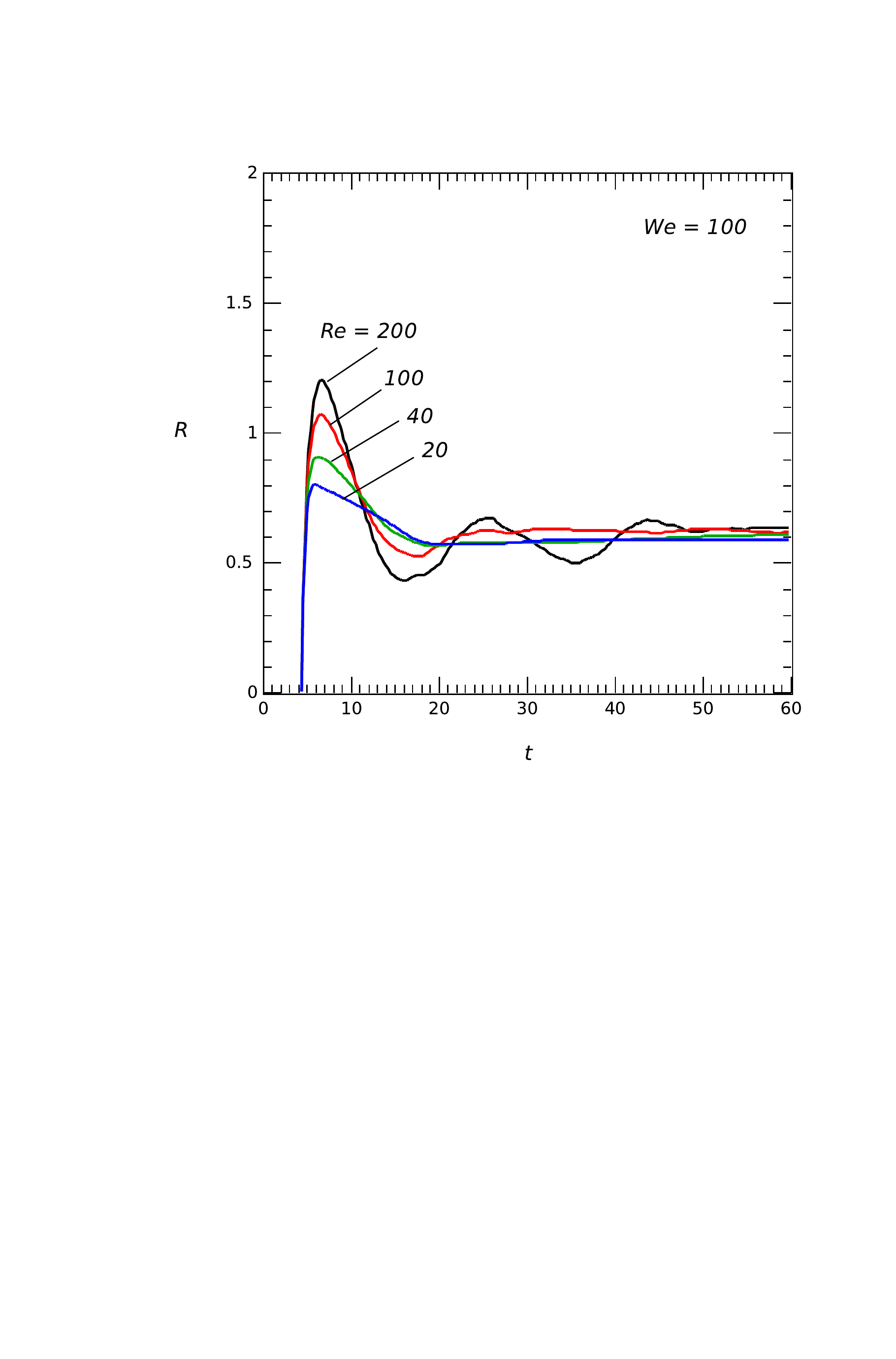}}
\caption{Plots of the center height $H$ 
and contact radius $R$ 
in units of $d$ versus time in units of $d/U$ ($= 0.25$ $\mu$s),
for droplets of $d = 5$ $\mu$m, $\rho_d = 2000$ kg m$^{-3}$,
$\sigma = 0.04$ N m$^{-1}$, 
with $\mu_d = 1$ cp ($Re = 200$ and $Ca = 0.5$), 
$2$ cp ($Re = 100$ and $Ca = 1$), 
$5$ cp ($Re = 40$ and $Ca = 2.5$),
and $10$ cp ($Re = 20$ and $Ca = 5$),
impacting solid surface at $U = 20$ m/s
for contact angle $\theta_0 = 90^o$.}
\label{fig:fig10}
\end{figure}

The computed values of maximum spread factor $\xi = 2 R_{max}$ 
for $We = 100$ at various $Re$
corresponding to various values of $\mu_d$ 
for droplets of $d = 5$ $\mu$m, $\rho_d = 2000$ kg m$^{-3}$,
$\sigma = 0.04$ N m$^{-1}$
with an impact velocity of $U = 20$ m/s
are given in table~\ref{t2},
along with that predicted by
(\ref{spread-factor}) and 
(\ref{spread-factor2}).
Again, the agreement between 
the present computations and 
either
(\ref{spread-factor}) or 
(\ref{spread-factor2})
is quite reasonable.
It should be noted that
the value of $\xi$ 
for the case of $\mu_d = 100$ cp ($Re = 2$)
is actually smaller than that at capillary equilibrium 
for $90^o$ contact angle (i.e., $1.26$).  
The $\xi$ in this case may not be literally regarded as
the ``maximum spread factor''.  
What is given here is actually the peak value of the spread factor
$2 R$ marking the end of spreading phase.  
In this case, the contact radius $R$ decreases slightly for a while 
from its peak value
in the relaxation phase, and then slowly increases toward 
the capillary equilibrium value $0.63$.

\begin{table}
\caption{\label{t2}Comparison of the present computed values of 
the maximum spread factor $\xi$
with that predicted by (\ref{spread-factor}) at $We = 100$, 
for droplets of $d = 5$ $\mu$m, $\rho_d = 2000$ kg m$^{-3}$,
$\sigma = 0.04$ N m$^{-1}$
with various values of viscosity $\mu_d$, 
when impacting solid surface at $U = 20$ m/s
for contact angle $\theta_0 = 90^o$ (with $\theta_A = 95^o$
and $\theta_R = 85^o$).}
%\begin{center}
\begin{indented}
\item[]\begin{tabular*}{0.75\textwidth}{@{\extracolsep{\fill}} c c c c c c c}
%\hline
\br
\\
$\mu_d$ (cp) & $Re$ & $Ca$ & $Oh$ & $\xi$ & Eq. (\ref{spread-factor}) & Eq. (\ref{spread-factor2}) \\
%\hline
\mr
1 & 200 & 0.5 & 0.05 & 2.415 & 2.154 & 2.594 \\
2 & 100  &  1 & 0.1 & 2.149 & 1.920 & 2.291 \\
5 & 40 & 2.5 & 0.25 & 1.819 & 1.649 & 1.938 \\
10 & 20 & 5 & 0.5 & 1.606 & 1.470 & 1.705 \\
100 & 2 & 50 & 5 & 1.143 & 1.003 & 1.103 \\
%\hline
\br
\end{tabular*}
\end{indented}
%\end{center}
\end{table}

To check the validity of 
(\ref{spread-factor}) and
(\ref{spread-factor2}) 
for contact angles
other than $\theta_0 = 90^o$, computations of 
cases for $\theta_0 = 45^o$ and 
$135^o$ (while other parameters remain 
unchanged from those in table 2) are also performed.
The results show that $\xi = 2.647$ and $2.272$
for $\mu_d = 1$ cp ($Re = 200$), 
$2.293$ and $2.064$ for $2$ cp ($Re = 100$), 
$1.905$ and $1.796$ for $5$ cp ($Re = 40$), 
$1.632$ and $1.596$ for $10$ cp ($Re = 20$), and 
$1.143$ and $1.143$ for $100$ cp ($Re = 2$).
A trend seems
to exist indicating a diminishing difference between the values $\xi$ 
for $\theta_0 = 45^o$ and $135^o$ with increasing $Oh$ 
which is a measure of relative strength of the viscosity effect.
In general, the computed values of $\xi$ (at $We = 100$)
are insensitive to the contact angle variations,
as consistent with the experimental findings of 
\cite{scheller1995} 
and reasoning of 
\cite{rioboo2002} for insignificant influence of contact angle to 
$\xi$.

However, the dynamics of free surface flow after initial spreading
can be quite sensitive to the contact angle 
when $We = 100$, similar to that shown for $We = 2500$.
A droplet with $\theta_0 = 45^o$ recedes very slowly whereas 
the contact line of that with $\theta_0 = 135^o$ moves rapidly 
during receding such that bouncing can occur.
Noteworthy here is that 
the same droplet of $d = 5$ $\mu$m, $\rho_d = 2000$ kg m$^{-3}$,
$\sigma = 0.04$ N m$^{-1}$
with viscosity $\mu_d = 5$, 
($Oh = 0.25$) and $\theta_0 = 135^o$ 
would bounce (i.e., detach) from substrate 
at $t = 26.8$
for lower impact velocity $U = 20$ m/s ($We = 100$) but
remain attached to the substrate at $U = 100$ m/s ($We = 2500$,
as indicated in figure 8).
This appears to be consistent with the trend shown by 
\cite{durickovic2005} 
for water drop impact on a solid surface and by 
\cite{law2015} 
in a general description of impact dynamics of droplets 
that bouncing 
is expected at lower $We$, 
due to large free surface deformation 
along the squeezing gas gap with reduced impact inertia, 
while merging-absorption at higher $We$,
due to sufficient impact inertia.

\subsection{Cases of $We = 5$}
\label{We5}
For a relatively small droplet of $d = 1$ $\mu$m, 
$\rho_d = 1000$ kg m$^{-3}$,
$\mu_d = 1$ cp, 
and $\sigma = 0.08$ N m$^{-1}$
with an impact velocity of $20$ m/s,
the values of $Re$ and $Ca$ become $20$ and $0.25$ such that
$We = 5$ and $Oh = 0.1118$.
For the same values of $Re$ and $Ca$ (as well as $We$ and $Oh$),
we can compute solutions with the same (dimensional) 
OpenFOAM mesh for 
$d = 5$ $\mu$m, 
$\rho_d = 200$ kg m$^{-3}$,
$\mu_d = 1$ cp, 
and $\sigma = 0.08$ N m$^{-1}$
with an impact velocity of $20$ m/s.

Figure 11 shows the variations of droplet surface profile 
with time for $We = 5$ and $Oh = 0.1118$.
In contrast to figures 2---4 for $We = 2500$ and $Oh = 0.05$,
the droplet surface in figure 11 does not form a commonly observed
thin lamella with a bulged rim at the end of spreading phase 
($t = 5.6$) due to lack of impact momentum.
The maximum contact radius at $t = 5.6$ is $0.6995$ 
while the center height reaches its minimum value of $0.4571$,
(which is much larger than $0.0266$ 
in subsection 3.1 
for $We = 2500$ and $Oh = 0.05$).
Following the end of spreading phase, the contact radius 
recedes and then oscillates with rather small amplitudes. 
Increasing liquid viscosity $\mu_d$,  
for a droplet of $d = 5$ $\mu$m,
$\rho_d = 200$ kg m$^{-3}$,
and $\sigma = 0.08$ N m$^{-1}$
with an impact velocity of $20$ m/s,
further reduces the magnitude of dynamics of free surface variations.
The value of contact radius $R$ could not 
even reach its capillary equilibrium value 
during the spreading phase for $\mu_d \ge 5$ cp
($Oh \ge 0.5590$);
rather it slowly creeps toward $0.63$ in the lengthy
wetting equilibrium phase. 
However, even for the case of $\mu_d = 10$ ($Oh = 1.1180$)
the center height $H$ still exhibits 
noticeable oscillatory motions, because 
of weaker viscous damping effect 
in the thick lamella away from the solid wall. 

If the liquid viscosity $\mu_d$ is increased to $10$ cp 
($Re = 2$ and $Ca = 2.5$),
the receding and oscillation phase disappears,
with the contact radius $R$ increase with time $t$ monotonically
as shown in figure 12.
However, there still seems to be a spreading phase
corresponding to a rapid increase of $R$, i.e., with 
a relatively large $dR/dt$, followed by a relaxation phase
with diminishing $dR/dt$ toward capillary 
equilibrium $R \approx 0.63$.
Because there are no local extrema for a peak of $R$,
the end of spreading phase cannot be clearly defined. 
The fact that 
the profile at $t = 10.0$ with a 
monotonically increasing contact radius $R$
has a center height $H$ slightly 
greater than that at $t = 5.6$ around the end of spreading phase,
indicating an oscillating free surface shape. 

\begin{figure}%[htb]
\centering
{\includegraphics[clip=true,scale=0.68,viewport=15 360 575 710]{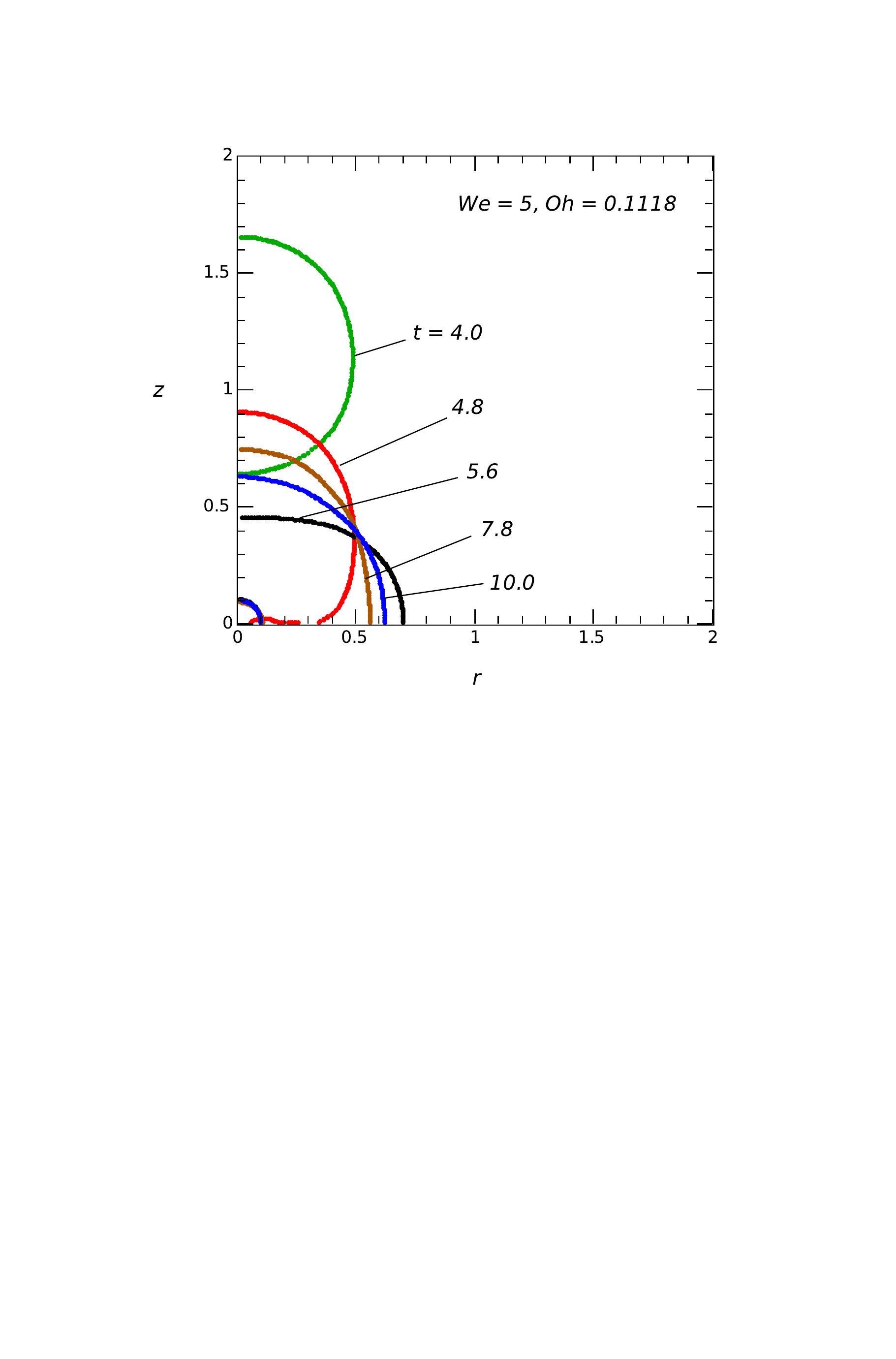}}
\caption{As in figure 2 but
for $\rho_d = 200$ kg m$^{-3}$, $\mu_d = 1$ cp, and 
$\sigma = 0.08$ N m$^{-1}$
($Re = 20$ and $Ca = 0.25$, or $We = 5$ and $Oh = 0.1118$)
with $\theta_0 = 90^o$
at $t = 4.0$, $4.8$, $5.6$, 
$7.8$,
and $10.0$.}
\label{fig:fig11}
\end{figure}

\begin{figure}%[htb]
\centering
{\includegraphics[clip=true,scale=0.68,viewport=15 360 575 710]{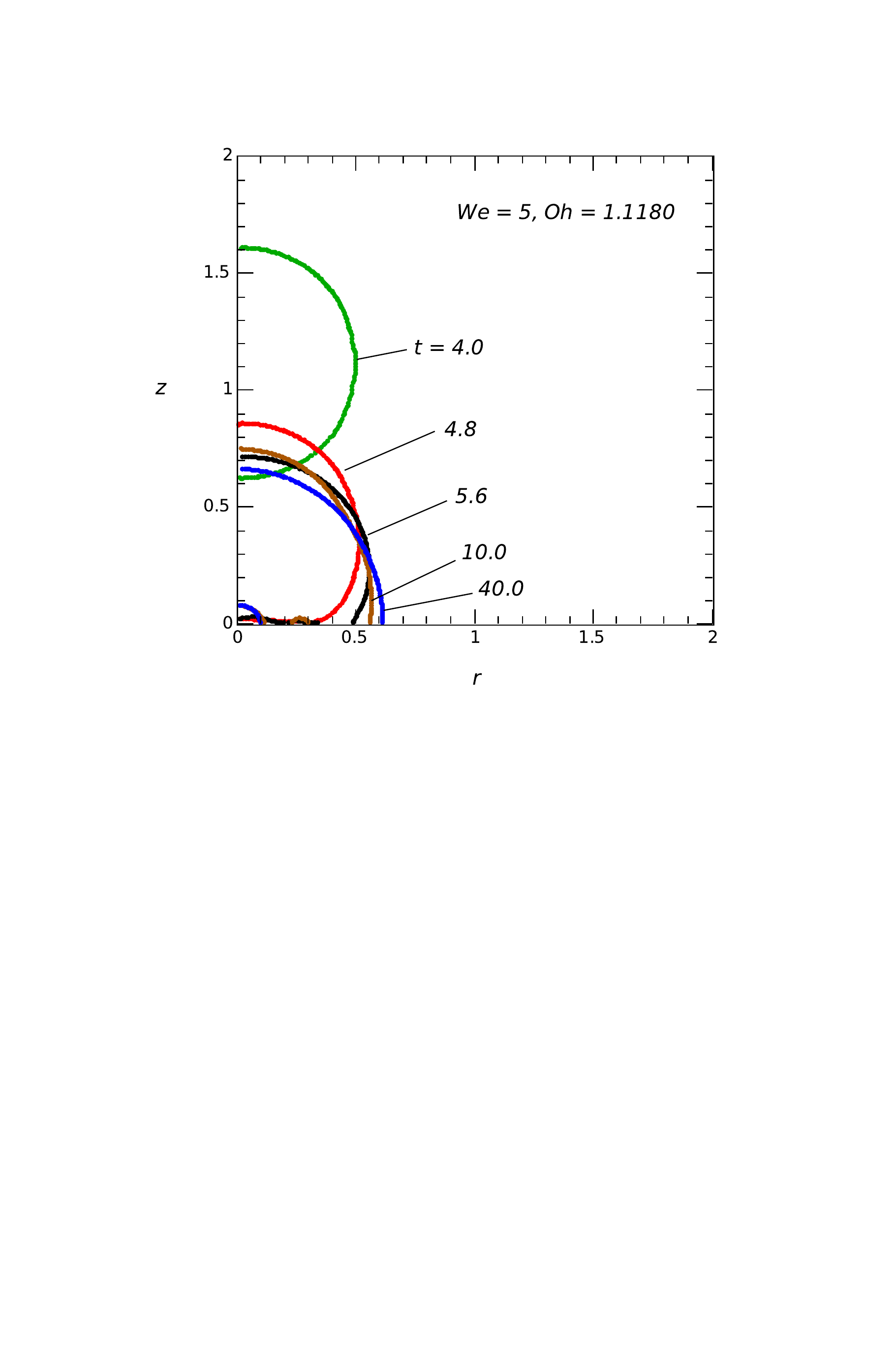}}
\caption{As in figure 11 but
for $\mu_d = 10$ cp 
($Re = 2$ and $Ca = 2.5$, or $We = 5$ and $Oh = 1.1180$)
at $t = 4.0$, $4.8$, $5.6$, 
$10.0$,
and $40.0$.}
\label{fig:fig12}
\end{figure}

Table~\ref{t3} shows
the computed values of maximum spread factor $\xi = 2 R_{max}$ 
for $We = 5$ at various $Re$
corresponding to various values of $\mu_d$ 
for droplets of $d = 5$ $\mu$m, $\rho_d = 200$ kg m$^{-3}$,
$\sigma = 0.08$ N m$^{-1}$
with an impact velocity of $U = 20$ m/s,
along with that predicted by
(\ref{spread-factor}) and 
(\ref{spread-factor2}).
Again, the agreement between 
the present computations and 
either
(\ref{spread-factor}) or 
especially (\ref{spread-factor2})
is still reasonable. 
For the cases of $Re \le 4$, 
the value of $\xi$ is taken as the (normalized) contact diameter 
at the end of spreading phase rather than literally the
maximum contact diameter.
However, the end of spreading phase may not be clearly defined.
In the case of $Re = 4$, a local extremum (or peak) of $R$ exists; 
so such a local peak value of $R$ is used 
as $R_{max}$ for calculating $\xi$. 
But in the cases of $Re = 2$ and $0.2$, $R(t)$ 
increases monotonically without local extrema;
the end of spreading phase can only be estimated based on
the slope change.  Therefore, the estimated values 
of $\xi$ in table~\ref{t3} 
are marked by an approximation sign ($\sim$).

\begin{table}
\caption{\label{t3}Comparison of the present computed values of 
the maximum spread factor $\xi$
with that predicted by (\ref{spread-factor}) at $We = 5$, 
for droplets of $d = 5$ $\mu$m, $\rho_d = 200$ kg m$^{-3}$, 
(or equivalently $d = 1$ $\mu$m, $\rho_d = 1000$ kg m$^{-3}$), 
$\sigma = 0.08$ N m$^{-1}$
with various values of viscosity $\mu_d$, 
when impacting solid surface at $U = 20$ m/s
for contact angle $\theta_0 = 90^o$ (with $\theta_A = 95^o$
and $\theta_R = 85^o$).}
%\begin{center}
\begin{indented}
\item[]\begin{tabular*}{0.75\textwidth}{@{\extracolsep{\fill}} c c c c c c c }
%\hline
\br
\\
$\mu_d$ (cp) & $Re$ & $Ca$ & $Oh$ & $\xi$ & Eq. (\ref{spread-factor}) & Eq. (\ref{spread-factor2}) \\
%\hline 
\mr
1 & 20 & 0.25 & 0.1118 & 1.399 & 1.146 & 1.302 \\ 
2 & 10  &  0.5 & 0.2236 & 1.285 & 1.022 & 1.192 \\ 
5 & 4 & 1.25 & 0.5590 & $\sim 1.120$ & 0.878 & 1.047 \\
10 & 2 & 2.5 & 1.1180 & $\sim 1.050$ & 0.782 & 0.942 \\
100 & 0.2 & 25 & 11.1803 & $\sim 0.835$ & 0.534 & 0.643 \\
\br
%\hline
\end{tabular*}
%\end{center}
\end{indented}
\end{table}

If computations are performed for  
cases of $We = 5$ with $\theta_0$ other than $90^o$ 
(while other parameters remain 
unchanged from those in table~\ref{t3}),
the results for the value of $\xi$ seems to still be
reasonably close to those for $\theta_0 = 90^o$, 
especially when the liquid viscosity $\mu_d$ is large.
For example,
with $\theta_0 = 135^o$ for $\mu_d = 5$ cp ($Re = 4$) and
$10$ cp ($Re = 2$)
a local extremum appears with
the peak value of contact radius 
$R_{max} = 0.503$ or $\xi = 1.006$ and 
$R_{max} = 0.469$ or $\xi = 0.938$, 
which are especially close to predicted values of 
$1.047$ and $0.942$ by
(\ref{spread-factor2}).
The value of $\xi$ remains the same at $0.835$ for the case of 
$\mu_d = 100$ cp at $\theta_0 = 135^o$.
But for $\mu_d = 1$ cp ($Re = 20$) and $\mu_d = 2$ cp ($Re = 10$),
the values of $R_{max}$ for $\theta_0 = 135^o$ 
become $0.5985$ or $\xi = 1.1970$
and $0.5558$ or $\xi = 1.1116$,
which are more than $10\%$ off
the corresponding values in table 3 for $\theta_0 = 90^o$.
Similarly with $\theta_0 = 45^o$ and $\mu_d = 1$ ($Ca = 0.25$),
the computed $\xi = 1.656$ is $\sim 18\%$ off $1.399$ in Table 3. 
Hence, the effect of contact angle $\theta_0$ 
on maximum spread factor $\xi$ seems to become more 
noticeable at small values of $We$ (e.g., $We < 10$),
especialy when $\mu_d$ (or $Ca$) is small.

As shown with the cases of $We = 2500$ and $100$,
bouncing tends to occur for 
droplet impact onto a hydrophobic surface.
However, for $We = 5$ with $\theta_0 = 135^o$, 
droplet bouncing after impact only happens when 
the liquid viscosity is very low, e.g., $\mu_d = 1$ cp
($Re = 20$).
Thus, 
the simple trend of bouncing at relatively smaller $We$  
\citep[e.g.,][]{durickovic2005, law2015} 
may not be general enough to cover all cases.
For example, the obvious effect of liquid viscosity on
the likelihood of bouncing illustrated in figure 8 
cannot be accounted for with the Weber number alone.

\section{Discussion}\label{discussion}
Generally speaking, 
(\ref{spread-factor}) and
(\ref{spread-factor2}) 
seems to describe the maximum spread factor $\xi$ fairly accurately
over a wide range of parameters for 
high-speed microdroplet impact.
with 
(\ref{spread-factor2}) being slightly more favorable in comparison
with the present computational results.
Noteworthy here is that different empirical, semiempirical formulas 
were many as proposed in the literature, with substantial discrepancies 
among each other 
\citep[as illustrated by][]{perelaer2009,
ravi2010, visser2012, visser2015}.
After comparing with several of the available formulas, 
(\ref{spread-factor}) and (\ref{spread-factor2}) 
are selected because the agreement between them and with 
the presently computed results appear to be quite consistent
across the ranges of parameters relevant to the 
Aerosol Jet$^{\circledR}$
technology. 
However, neither 
(\ref{spread-factor}) nor 
(\ref{spread-factor2}) 
explicitly accounts for the contact angle effect,
which tends to become more noticeable with reduced $We$ 
and small $Ca$ (e.g., $We = 5$ and $Ca < 1$).
Unlike
(\ref{spread-factor}) and
(\ref{spread-factor2}),
the maximum spread factor formula derived by 
\cite{pasandideh1996} based on energy balance 
actually contains the contact angle $\theta_0$ as
\begin{equation}\label{spread-factor3}
\xi = \sqrt{\frac{12 + We}{3 (1 - \cos\theta_0) + 4 \sqrt{Ca \, We}}} 
\quad .
\end{equation}
Despite its considerable discrepancy from the 
computed $\xi$ values in the present work,
(\ref{spread-factor3}) indeed suggests that
the contact angle effect should diminish when
the value of $Ca \times We$ becomes close to $100$ or greater.
It also qualitatively predicts the trend of decreasing 
contact angle effect on $\xi$ with increasing $Ca$ 
(or viscosity $\mu_d$) as consistent with the present results. 
In view of the general accuracy of  
(\ref{spread-factor2}) and reasonable account for the trend of
contact angle effect with 
(\ref{spread-factor3}),
a straightforward combination of the two yields 
\begin{equation}\label{spread-factor4}
\xi = \left(Re^{1/5} - 0.35 \frac{Re^{2/5}}{\sqrt{We}}\right) 
\sqrt{\frac{3 + 4 \sqrt{Ca \, We}}
{3 (1 - \cos\theta_0) + 4 \sqrt{Ca \, We}}} 
\quad .
\end{equation}
For example, in the case of 
a droplet with $d = 1$ $\mu$m, $\rho_d = 1000$ kg m$^{-3}$), 
$\sigma = 0.08$ N m$^{-1}$, and $\mu_d = 1$ cp impacting the substrate
at $U = 20$ m/s (i.e., $We = 5$ and $Ca = 0.25$) and $\theta_0 = 135^o$, 
the computed $\xi = 1.197$ 
and that calculated with (\ref{spread-factor4}) is $1.149$ 
whereas with (\ref{spread-factor2}) $1.302$.
If $\mu_d$ is increased to $2$ cp (i.e., $We = 5$ and $Ca = 0.5$),
the computed $\xi = 1.112$ 
and that calculated with (\ref{spread-factor4}) is $1.076$ 
whereas with (\ref{spread-factor2}) $1.192$.
With $\theta_0 = 45^o$ for $\mu_d = 1$ and $2$ cp ($Ca = 0.25$ and $0.5$),
the computed $\xi$ are $1.656$ and $1.379$ 
while that predicted by (\ref{spread-factor4})
are $1.539$ and $1.356$, 
much improved from $1.302$ and $1.192$ by (\ref{spread-factor2}).
Thus, (\ref{spread-factor4}) can be 
a useful formula with improved accuracy for the range of parameters of 
practical interest to 
applications with the
Aerosol Jet$^{\circledR}$
direct-write technology.

While the maximum spread factor $\xi$ provides a 
practically useful correlation between the Feret diameter of deposited 
individual droplet on a dry substrate and the diameter of 
corresponding droplet before impact, 
other dynamic outcomes of droplet impact can be relevant to
Aerosol Jet$^{\circledR}$
ink deposition, too.
Although not computed with the present axisymmetric model,
the splashing phenomenon usually observed in droplet impact
with large $We$ and $Re$ 
is also of great importance to 
Aerosol Jet$^{\circledR}$
printing for being a possible source of undesirable oversprays
and uncontrolled satellites.
Historically,
the first study of splashing after droplet impact was 
carried out by \cite{worthington1876}.
Splashing of large milk and mercury droplets onto smooth glass plates
was observed and the corresponding fingering patterns were sketched,
with the number of fingers increasing with both droplet size and 
fall height being noted.
Several investigations on surface roughness effect on
splashing behavior suggested that 
splashing at atmospheric pressure only occurs when
\begin{equation}\label{splash-parameter}
K \equiv \frac{We}{Oh^{2/5}} > K_s 
\quad \mbox{ with } \quad
K_s = 649 + \frac{3.76}{R_a^{0.63}}
\quad ,
\end{equation}
where $R_a$ denotes the nondimensional roughness parameter 
in units of $d$ 
\citep{stow1981, mundo1995, cossali1997, yarin2006}.
Thus, $K_s \to 649$ as $R_a \to \infty$ for a very rough surface,
whereas $K_s$ increases to infinity
as $R_a \to 0$ for an extremely smooth surface.
However, other forms for splashing criteria had also been proposed
in the literature
\citep[e.g.,][]{moreira2010, mandre2012, stevens2014}, 
but little agreement had been shown among different proposed criteria 
which often contradict one another \citep{visser2015}.
While careful examining the validity of each proposed criterion
is out of scope of the present work,
(\ref{splash-parameter})
may be used as a tentative reference for a brief discussion here.

In the case of 
Aerosol Jet$^{\circledR}$
ink droplets with $d = 5$ $\mu$m,
$R_a = 0.1$ (which leads to $K_s = 665$, only slightly greater
than $649$) 
corresponds to a roughness length scale
($0.5$ $\mu$m) 
around the wavelengths of visible light 
which is usually considered as a fairly smooth surface 
with most of realistic substrate surfaces.
For cases with $U = 100$ m/s ($We = 2500$), as those in table~\ref{t1},
the values of $K$ are all exceeding $649$ (or $665$), 
ranging from $1313$ for $Re = 10$ to
$8286$ for $Re = 1000$. 
Thus, when operating at a very high jet speed (e.g., $U = 100$ m/s)
under atmospheric pressure, the
Aerosol Jet$^{\circledR}$
ink droplets of $d = 5$ $\mu$m are expected to disintegrate
as a consequence of splashing after impacting the substrate.
If the ink droplet $d$ is 
reduced to $2$ $\mu$m with $\mu_d = 100$ cp at the same $U$,
the value of $K$ can become $437$ ($< 649$).
Even for an ink droplet of $d = 1$ $\mu$m with $\mu_d = 5$ cp
(for $\rho_d = 2000$ kg m$^{-3}$ and $U = 100$ m/s),
the value of $K$ is $631$, barely below the 
reference splashing threshold value $649$.
Thus, to avoid ink droplet splashing upon deposition in
Aerosol Jet$^{\circledR}$
printing with high-speed jet,
it is preferrable to keep the droplet size small and 
viscosity high (which may be accomplished by 
enabling effective in-flight mist solvent evaporation).

\section{Concluding remarks}\label{conclusion}
In view of the challenges with required high spatial and temporal
resolutions for experimentally analyzing 
the ink droplet deposition behavior during
Aerosol Jet$^{\circledR}$
printing (with microdroplets of diameters from $1$ to $5$ $\mu$m 
and impact velocity from $20$ to $100$ m/s),
numerical solutions for high-speed microdroplet impact onto 
a smooth solid surface are computed in the present work using 
the {\it interFoam} VoF solver of the
OpenFOAM$^{\circledR}$
CFD package.
For simplicity and computational efficiency,
the free-surface fluid dynamics problem is assumed to be
axisymmetric with incompressible flow.
The computed results illustrate 
droplet impact dynamics with lamella shape evolution 
throughout the spreading, receding-relaxation, and
wetting equilibrium phases, consistent with what have been 
observed and described in various previous studies.
This fact agrees with the conclusions of \cite{visser2015} that 
the basic droplet impact behavior is scale-invariant;
in other words, experiments with larger droplets at 
the same nondimensional parameter values should be 
able to describe the phenomena with much smaller droplets.
When the droplet viscosity is relatively low,
significant oscillations in the free-surface flow 
can be observed.
But the free surface oscillatory motion 
seems to diminish 
as the droplet viscosity $\mu_d$ becomes relatively high.
The border line between periodic free surface oscillations 
and aperiodic creeping to capillary equilibrium free surface shape 
after impact spreading appears at
Ohnesorge number $Oh$ 
($\equiv \mu_d/\sqrt{\rho_d \sigma d}$) about $0.25$.
Understanding the controlling factors for free surface oscillations
can be important for 
Aerosol Jet$^{\circledR}$
ink formulation and process recipe development.
 
The computed results show that substrate surface properties such as
the contact angle can drastically influence the dynamics of 
free surface deformation after the spreading phase.
For example, 
droplet bouncing (i.e., rebound) is prompted with  
large contact angles at solid surface 
\citep[i.e., hydrophobic surface, consistent with 
findings reported in the literature, cf.][]{rioboo2001, durickovic2005}, 
but its likelihood can be reduced by
increasing the droplet viscosity 
due to enhanced kinetic energy dissipation.
At some intermediate viscosity values,
reattachmet of the bouncing droplet to 
the solid surface can be observed within a short time.
When using a high-speed jet flow to direct the ink droplet deposition in
Aerosol Jet$^{\circledR}$ 
printing,
droplet bouncing after impact on substrate is generally undesirable for 
causing unintended ink placement such as 
``satellite'', ``overspray'', etc. 

Special attention has been paid to the value of
maximum spread factor $\xi$,
which can be accurated determined from 
the numerical solutions.
Given substantial discrepancies
among different correlations by many authors in the 
literature, comparisons with the presently computed 
$\xi$ have been performed to construct a 
useful formula with reasonable accuracy. 
For the range of parameters of practical interest to the 
Aerosol Jet$^{\circledR}$
printing,
the values of computed $\xi$ agree quite well with
the empirical correlation of \cite{scheller1995}
based mostly on experimental data
and the semiempirical relation 
proposed by \cite{roisman2009}
based on an analytical theory for inertia dominated situations
(with a slight modification of the coefficient values).
Majority of the computed cases show insignificant variations of $\xi$ 
with changes of contact angle $\theta_0$, as expected when 
dynamics in the spreading phase is dominated by inertial effect.
The weak dependence of $\xi$ on contact angle $\theta_0$,
especially becoming more noticeable at relatively small $Ca$ and $We$,
can be accounted for with 
a straightforward combination of 
the formula of \cite{pasandideh1996} 
and that of \cite{roisman2009}. 
The resulting maximum spread factor formula can be used for 
first-order evaluations of deposited ink droplet size during
Aerosol Jet$^{\circledR}$
technology development.

\section*{Acknowledgment}
The author would like to thank John Lees for encouragement and 
guidance, and Dr. Mike Renn for insightful discussion.

%\bibliographystyle{model4-names}
%\bibliography{<your-bib-database>}

\section*{References}
\begin{harvard}
\bibitem[Antonini et al.(2012)]{antonini2012} 
Antonini, C., Amirfazli, A., and Marengo, M., 
2012 Drop impact and wettability: From hydrophilic to 
superhydrophobic surfaces,
Phys. Fluids 24, 102104

\bibitem[Attane et al.(2007)]{attane2007} Attane, P.,
Girard, F., and Morin, V., 2007 An energy balance approach of
the dynamics of drop impact on a solid surfaces.
Phys. Fluids 19, 012101 

\bibitem[Berberovic et al.(2009)]{berberovic2009}
Berberovic, E., Van Hinsber, N. P., Jakirlic, S., Roisman, I. V.,
Tropea, C., 2009.
Drop impact onto a liquid layer of finite thickness: Dynamics of the
cavity evolution.
{\it Phys. Rev. E} 79(3), 036306.

\bibitem[Bussmann et al.(1999)]{bussmann1999} Bussmann, M.,
Mostaghimi, J., Chandra, S. 1999 On a three-dimensional 
volume tracking model of droplet impact.
Phys. Fluids 11(6), 1406--1417

\bibitem[Bussmann et al.(2000)]{bussmann2000} Bussmann, M.,
Chandra, S., Mostaghimi, J. 2000 Modeling the splash of
a droplet impacting a solid surface.
Phys. Fluids 12(12), 3121--3132

\bibitem[Chandra and Avedisian(1991)]{chandra1991} Chandra, S.
and Avedisian, C. T. 1991
On the collision of a droplet with a solid surface.
Proc. R. Soc. London A432, 13--41

\bibitem[Christenson et al.(2011)]{christenson2011}
Christenson, K.K., Paulsen, J.A., Renn, M.J., McDonald, K., and Bourassa, J. 2011
Direct printing of circuit boards using
Aerosol Jet$^{\circledR}$.
Proc. NIP 27 Digital Fabric., 433--436

\bibitem[Cossali et al(1997)]{cossali1997} 
Cossali, G. E., Coghe, A., and Marengo, M., 
1997 The impact of a single drop on a wetted sold surface. 
Exp. Fluids 22, 463--472

\bibitem[Deshpande et al.(2012)]{deshpande2012} 
Deshpande, S. S., Lakshman, A., and Trujillo, M. F., 
2012 Evaluating the performance of the two-phase flow solver
interFoam. 
Comput. Sci. Discov. 5, 04016 

\bibitem[Dinc and Gray(2012)]{dinc2012} 
Dinc, M. and Gray, D. D., 
2012 Drop impact on a wet surface: computational investigation of  
gravity and drop shape. 
Advance in Fluid Mechanics and Heat \& Mass Transfer
(ISBN: 978-1-61804-114-2), pp 374--379 

\bibitem[Durickovic and Varland(2005)]{durickovic2005} 
Durickovic, B. and Varland, K., 
2005 Between bouncing and splashing: water drops on a solid surface. 
Applied Mathematics thesis, USA; University of Arizona

\bibitem[Feng(2010)]{feng2010} Feng, J. Q.
2010 A deformable liquid drop falling through a quiescent
gas at terminal velocoty.
J. Fluid Mech. 658, 438--462

\bibitem[Feng(2015)]{feng2015} Feng, J. Q.
2015 Sessile drop deformations under an impinging jet.
Theor. Comput. Fluid Dyn. 29, 277--290

\bibitem[Ford and Furmidge(1967)]{ford1967} Ford, R. E. 
and Furmidge, C. G. L. 1967 Impact and spreading of 
spray drops on foliar surfaces. Wetting, Soc. Chem. 
Industry Monograph, 417--432

\bibitem[Foote(1974)]{foote1974} Foote, G. B. 1974 The water drop
rebound problem: dynamics of collision. J. Atmos. Sci. 32, 390--402

\bibitem[German and Bertola(2009)]{german2009} German, G.,
and Bertola, V., 2009 Review of drop impact models and
validation with high-viscosity Newtonian fluids.
Atom. Sprays 19(8), 787--807 

\bibitem[Gopala and van Wachem(2008)]{gopala2008} Gopala, V. R.
and van Wachem, B. G. M., 2008 Volume of fluid methos for 
immiscible-fluid and free-surface flows.
Chem. Eng. J. 141(1), 204--221 

\bibitem[Gupta and Kumar(2010)]{gupta2010} 
Gupta, A. and Kumar, R., 
2010 Droplet impingement and breakup on a dry surface. 
Comput. Fluids 39, 1696--1703

\bibitem[Healy et al.(1996)]{healy1996} Healy, W. M.,
Hartely, J. G., Abdel-Khalik, S. I. 1996 Comparison between
theoretical models and experimental data for the spreading
of liquid droplets impacting a solid surface.
Int. J. Heat Mass Transfer 39, 3079--3082

\bibitem[Hedges et al.(2007)]{hedges2007}
Hedges, M., King, B. and Renn, M. 2007 Direct writing for  
advanced electronics packaging.
www.onboard-technology.com/pdf\_giugno2007/060706.pdf

\bibitem[Hoang et al.(2013)]{hoang2013} Hoang, D. A.,
van Steijn, V., Portela, L. M., Kreutzer, M. T.,
and Kleijn, C. R. 2013 Benchmark numerical simulations of segmented 
two-phase flows in microchannels using the volume of fllluid method.
Comput. Fluids 86, 28--36

\bibitem[Kahn(2007)]{kahn2007}
Kahn, B.E. 2007 The M$^3$D aerosol jet system,
an alternative to inkjet printing for printed electronics,
Organic and Printed Electronics, 1, 14--17

\bibitem[Landau and Lifshitz(1959)]{landau1959} 
Landau, L. D. and Lifshitz, E. M. 
1959 Fluid Mechanics. 
Addison-Wesley  

\bibitem[Law(2015)]{law2015} 
Law, C. K. 
2015 Impact dynamics of droplets and jets. 
ICLASS2015, 13th Int. Conf. Liquid Atom. Spray Syst. Tainan, Taiwan  

\bibitem[Linder et al.(2013)]{linder2013} 
Linder, N., Roisman, I. V., Marschall, H., and Tropea,  C. 
2013 Numerical simulations of pinning droplets. 
ILASS--Europe 2013, 25th Europ. Conf. Liquid Atom. Spray Syst. 
Chania, Greece  

\bibitem[Mandre and Brenner(2012)]{mandre2012} 
Mandre, S. and Brenner, M. P., 
2012 The mechanism of a splash on a dry solid surface.
J. Fluid Mech. 690, 148--172

\bibitem[Moreira et al(2010)]{moreira2010} 
Moreira, A. L. N., Moita, A. S., and Panao, M. R., 
2010 Advances and challenges in explaining fuel spray impingement:
how much of single droplet impact research is useful? 
Prog. Energy Combust. Sci. 36, 554--580

\bibitem[Morgan(2013)]{morgan2013} 
Morgan, G. C. J., 
2013 Application of the interFoam VoF code to
coastal wave/structure interaction.
PhD thesis: University of Bath

\bibitem[Mundo et al.(1995)]{mundo1995} 
Mundo, C., Sommerfeld, M.. and Tropea, C., 
1995 Droplet-wall collisions: experimental studies  
of the deformation and breakup process.
Int. J. Multiphase Flow 21, 151--173

\bibitem[Pasandideh-Fard et al.(1996)]{pasandideh1996} 
Pasandideh-Fard, M., Qiao, Y. M., Chandra, S. and Mostaghimi, J.
1996 Capillary effects during droplet impact 
on a solid surface.
Phys. Fluids 8, 650--659

\bibitem[Paulsen et al.(2012)]{paulsen2012} 
Paulsen, J. A., Renn, M., Christenson, K., and Plourde, R. 
2012 Printing conformal electronics on 3D structures 
with Aerosol Jet technology.
In Future of Instrumentation International Workshop (FIIW)
doi: 10.1109/FIIW.2012.6378343

\bibitem[Perelaer et al.(2009)]{perelaer2009} Perelaer, J.,
Smith, P. J., van den Bosch, E., van Grootel, S. S. C.,
Ketelaars, P. H. J. M., and Schubert, U. S. 2009 
The spreading of inkjet-printed droplets with varying
polymer molar mass on a dry solid substrate.
Macromol. Chem. Phys. 210, 495--502 

\bibitem[Ravi et al.(2010)]{ravi2010} Ravi, V.,
Jog, M. A., and Manglik, R. M. 2010 Effects of interfacial and
viscous propertis of liquids on drop spread dynamics.
ILASS2010-142, 22nd Ann. Conf. Liquid Atom. Spray Syst. Cincinnati, OH  
\bibitem[Rein(1993)]{rein1993} Rein, M. 1993 Phenomena
of liquid drop impact on solid and liquid surfaces.
Fluid Dyn. Res. 12, 61--93

\bibitem[Renn(2006)]{renn2006} Renn, M. J. 
2006 Direct Write$^{TM}$ system.
US Patent 7,108,894 B2

\bibitem[Renn(2007)]{renn2007} Renn, M. J. 
2007 Direct Write$^{TM}$ system.
US Patent 7,270,844 B2

\bibitem[Renn et al.(2009)]{renn2009} Renn, M. J.,
King, B. H., Essien, M., Marquez, G. J., Giridharan, M. G., 
and Sheu, J.-C. 2009 
Apparatuses and methods for maskless mesoscale material deposition. 
US Patent 7,485,345 B2

\bibitem[Renn et al.(2010)]{renn2010} Renn, M., Essien, M.,
King, B. H., and Paulsen, J. A. 2010 
Aerodynamic jetting of aerosolized fluids 
for fabrication of passive structures.
US Patent 7,674,671 B2

\bibitem[Rioboo et al.(2001)]{rioboo2001} 
Rioboo, R., Tropea, C., and Marengo, M. 
2001 Outcomes from a drop impact on solid surfaces.
Atomiz. Sprays 11, 155--165

\bibitem[Rioboo et al.(2002)]{rioboo2002} 
Rioboo, R., Marengo, M., and Tropea, C. 
2002 Time evolution of liquid drop impact onto solid,
dry surface.
Exp. Fluids 33, 112--124

\bibitem[Roisman(2009)]{roisman2009}
Roisman, I. V. 2009
Inertia dominated drop collisions. II. An analytical solution
of the Navier-Stokes equations for a spreading viscous film. 
Phys. Fluids 21, 052104

\bibitem[Rusche(2002)]{rusche2002} Rusche, H.
2002 Computational fluid dynamics of dispersed  
two-phase flows at high phase fractions.
PhD thesis, University of London / Imperial College

\bibitem[Saha and Mitra(2009)]{saha2009} Saha, A. A.
and Mitra, S. K. 2009 Effect of dynamic contact angle in a  
volume of fluid (VoF) model for a microfluidic capillary flow.
J. Colloid Interf. Sci. 339, 461--480

\bibitem[Scheller and Bousfield(1995)]{scheller1995} Scheller, B. L.
and Bousfield, D. W. 1995 Newtonian drop impact 
with a solid surface.
AIChE J. 41, 1357--1367

\bibitem[Sikalo et al.(2002)]{sikalo2002} Sikalo, S.,
Marengo, M., Tropea, C., Ganic, E. N. 2002 Analysis of
impact of droplet on horizontal surfaces.
Exp. Therm. Fluid Sci. 25, 503--510

\bibitem[Stevens(2014)]{stevens2014} 
Stevens, C. S. 
2014 Scaling of the splash threshold for low viscosity fluids.  
Europhys. Lett. 106, 24001

\bibitem[Stow and Hadfield(1981)]{stow1981} 
Stow, C. D. and Hadfield, M. G. 
1981 An experimental investigation of fluid flow resulting from  
the impact of a water drop with an unyielding dry surface. 
Proc. R. Soc. Lond. A373, 419--441

\bibitem[Tanner(1979)]{tanner1979} 
Tanner, L. H. 
1979 The spreading of silicone oil drops on solid surfaces. 
J. Phys. D 12, 1473--1484

\bibitem[Toivakka(2003)]{toivakka2003} Toivakka, M.
2003 Numerical investigation of droplet impact spreading
in spray coating of paper.
Proceedings of TAPPI 8th Advanced Coating Fundamentals Symposium

\bibitem[Ubbink(2002)]{ubbink02}
Ubbink, H. 2002
Computational Fluid Dynamics of Dispersed Two-Phase Flows at
High Phase Fractions. PhD Thesis, Imperial College of Science,
Technology and Medicine, London

\bibitem[Visser et al.(2012)]{visser2012} Visser, C. W.,
Tragawa, Y., Sun, C.,and Lohse, D. 2012 Microdroplet impact
at very high velocity.
Soft Matter 8, 10732  

\bibitem[Visser et al.(2015)]{visser2015} Visser, C. W.,
Frommhold, P. E., Wildeman, S., Mettin, R. Lohse, D., and Sun, C. 
2015 Dynamics of high-speed micro-drop impact: numerical simulations
and experiments at frame-to-frame times below 100ns.
Soft Matter 11, 1708-1722  

\bibitem[Worthington(1876)]{worthington1876} 
Worthington, A. M. 
1876 On the forms assumed by drops of liquids 
falling vertically on a horizontal plate. 
Proc. R. Soc. Lond. 25, 261--271

\bibitem[Yarin(2006)]{yarin2006} Yarin, A. L.
2006 Drop impact dynamics: splashing, spreading, receding, bouncing...
Ann. Rev. Fluid Mech. 38, 159--192

\bibitem[Zollmer et al.(2006)]{zollmer2006}
Zollmer, V., Muller, M., Renn, M., Busse, M., Wirth, I., Godlinski, D.,
and Kardos, M. 2006 Printing with aerosols: A maskless deposition
technique allows high definition printing of a variety of
functional materials,
Euro. Coating J., 07-08, 46--55


\end{harvard}
%\begin{thebibliography}{00}

%\end{thebibliography}

\end{document}